\documentclass[12pt, oneside]{article}
\usepackage[utf8]{inputenc}
\usepackage{appendix}
\usepackage{geometry} 
\usepackage{amsmath} 					\usepackage{fancyhdr}			
\usepackage{graphicx}	
\usepackage{amssymb}
\usepackage{cite}
\usepackage{color,xspace}
\usepackage{extarrows}
\usepackage{mathtools}
\usepackage{comment}
\usepackage{braket}
\usepackage[skip=2pt,font=scriptsize]{caption}

\usepackage[margin=1cm]{caption}

\DeclareSymbolFont{extraup}{U}{zavm}{m}{n}
\DeclareMathSymbol{\vardiamond}{\mathalpha}{extraup}{87}
\usepackage{float}
\restylefloat{table}
\allowdisplaybreaks

\def\twomat[#1,#2][#3,#4]{\left( \begin{array}{cc} #1 & #2 \\ #3 & #4 \end{array} \right)}
\def\thv[#1,#2,#3]{\left( \begin{array}{c} #1 \\ #2 \\ #3 \end{array} \right)}
\def\twv[#1,#2]{\left( \begin{array}{c} #1 \\ #2 \end{array} \right)}

\title{The dilation and Scalar Weak Gravity Conjecture }

\date{}

\begin{document}

\begin{flushright}
\end{flushright}
\begin{center}

\vspace{1cm}
{\LARGE{\bf Dilatonic (Anti-)de Sitter  Black Holes and  Weak Gravity Conjecture }}

\vspace{1cm}

\large{\bf Karim Benakli$^\spadesuit$ \let\thefootnote\relax\footnote{$^\spadesuit$kbenakli@lpthe.jussieu.fr},
Carlo Branchina$^{\vardiamond}$ \let\thefootnote\relax\footnote{$^\vardiamond$cbranchina@lpthe.jussieu.fr}
and
Ga\"etan~Lafforgue-Marmet$^\clubsuit$ \footnote{$^\clubsuit$glm@lpthe.jussieu.fr}
 \\[5mm]}

{ \sl Laboratoire de Physique Th\'eorique et Hautes Energies (LPTHE),\\ UMR 7589,
Sorbonne Universit\'e et CNRS, 4 place Jussieu, 75252 Paris Cedex 05, France.}

\end{center}
\vspace{0.7cm}

\abstract{ 
 Einstein-Maxwell-dilaton theory  with non-trivial dilaton potential is known to admit asymptotically flat and (Anti-)de Sitter charged black hole solutions. We investigate the conditions for the presence of horizons as function of the parameters mass $M$, charge $Q$ and dilaton coupling strength $\alpha$. We observe that there is a value of $\alpha$ which separate two regions, one where the black hole is Reissner-Nordstr\"om-like from a region where it is Schwarzschild-like. We find that for de Sitter and small non-vanishing $\alpha$, the extremal case is not reached by the solution. We also discuss the attractive or repulsive nature of the leading long distance interaction between two such black holes, or a test particle and one black hole, from a world-line effective field theory point of view. Finally, we discuss possible modifications of the Weak Gravity Conjecture in the presence of both a dilatonic coupling and a cosmological constant.
}

\newpage

\tableofcontents

\setcounter{footnote}{0}

\section{Introduction }

While global symmetries seem fine in Quantum Field Theory (QFT), their existence is a no-go for any Quantum Theory of Gravity: global charges are not conserved when falling in black holes (see \cite{Banks:2010zn}). This illustrates the fact that not all model builder's ingredients are allowed in Nature: some consistent QFTs will never be derived from an ultraviolet (UV) theory that includes quantum gravity, such as String Theory. They fall in the Swampland, contrary to those that form the Landscape (see \cite{Vafa:2005ui}). Maybe the best-tested condition that discriminates between the two sets of theories is the Weak Gravity Conjecture \cite{ArkaniHamed:2006dz}. This requires, for an abelian $U(1)$ gauge symmetry, the presence of at least one state carrying a charge $Q$ bigger than its mass $M$, measured in Planck length units:
${M^2} < Q^2 $. Arguments based on black hole (BH) physics have allowed to extend this conjecture to either Einstein-Maxwell-dilaton theory in flat space-time \cite{Heidenreich:2015nta} or to de Sitter space-time \cite{Antoniadis:2020xso}. It was the aim of this work to put the two together: Einstein-Maxwell-dilaton theory with (Anti-) de Sitter ((A)dS) backgrounds.\\

The present work began then when we asked ourselves what happens to the de Sitter Weak Gravity Conjecture in the case of a dilatonic gauge coupling. In \cite{Antoniadis:2020xso}, the loci of black hole horizons were interpreted as the result of the competition between a repulsive electromagnetic energy density on one side, and gravity on the other, with attractive and repulsive contributions from the black hole mass and the cosmological constant, respectively. It was suggested that the parameter region where the electromagnetic contribution dominates defines simultaneously the absence of a black hole solution and the WGC conditions. It was comforting to see that in the limit of  vanishing cosmological constant one recovers the flat space-time result, in contrast with previous attempts \cite{Huang:2006hc,Montero:2019ekk} that provide complementary criteria for the consistency of the theory and therefore different proposals for a WGC. We were also interested in the dilaton as an example of scalar field that allows to probe the Scalar WGC \cite{Benakli:2020pkm,Palti:2017elp,Gonzalo:2020kke,Heidenreich:2019zkl,Gonzalo:2019gjp,Freivogel:2019mtr,Ellis:2020nnp,Calderon-Infante:2020dhm,Noumi:2021uuv}.\\  

An extension of the Reissner-Nordstr\"om de Sitter black hole solution \cite{Kottler,Ginsparg:1982rs,Romans:1991nq} to the case of Einstein-Maxwell-dilaton theory  was constructed in \cite{Gao:2004tu,Elvang:2007ba,Mignemi:2009ui}. This is a generalisation of the well known flat space-time solution of Gibbons-Maeda \cite{Gibbons:1987ps} and Garfinkle-Horowitz-Strominger \cite{Garfinkle:1990qj}. For the de Sitter background, we were not able to find in the literature a discussion on the conditions for the existence of horizons with a dilatonic coupling $\alpha \neq 0$. Some aspects of the asymptotically Anti-de Sitter metric were discussed for the case of AdS$_5$ in \cite{Elvang:2007ba} and for AdS$_4$ in some limits by \cite{Goto:2018iay}. Some properties of these solutions, as the photon spheres, were considered in \cite{Cvetic:2016bxi}. It is the main goal of this work to provide the missing comprehensive analysis. In the different cases, the WGC states will then be considered to be contained within the parametric regions complementary to those where black holes exist.\\

The paper is organized as follows. Section 2 presents the black hole solution and some formulae generic to all values of $\alpha$. A very brief review of the asymptotically flat space-time is given in Section 3 for completeness and comparison. The horizons of charged dilatonic de Sitter black hole are described in Section 4. The Anti-de Sitter case is studied in Section 5. Some thermodynamic quantities are computed in Section 6. The issue of attractive and repulsive forces, in the case of asymptotically flat space-time, are analyzed in Section 7. For the convenience of the readers, our results are summarized in Section 8 with our conclusions.   

\section{Einstein-Maxwell-dilaton Black Holes}

The Reissner-Nordstr\"om Black Holes are parametrized by their charge $\tilde Q$ and  mass $\tilde M$. It is useful to define the analog of such quantities in the so-called geometrized units: 
\begin{equation} 
M=\frac{\kappa^2\,\tilde M}{8\pi}, \qquad Q^2=\frac{\kappa^2 \tilde Q^2}{32\pi^2} \qquad \Rightarrow \qquad \frac{M^2}{Q^2}=\frac{\kappa^2}{2}\frac{\tilde M^2}{\tilde Q^2} ,
\end{equation}
with $\kappa^2={1}/{M_P^2}=8\pi G\equiv 8\pi$ and $G$ Newton's constant.
The absence of a naked singularity requires $Q\le M$.\\

In the following we consider the extension provided by the Einstein-Maxwell-dilaton action
\begin{equation}
\label{Einstein Maxwell dilaton deSitter action}
\mathcal S=\int d^4x \sqrt{-g}\frac{1}{2\kappa^2}\left(R-2\left(\partial \phi\right)^2-e^{-2\alpha\phi}F^2 -V(\phi) \right),
\end{equation}
where $F_{\mu\nu}$ is the field strength tensor related to the massless gauge field $A_\mu$. 
For $V(\phi) =0$, the values $\alpha=1$ and $\alpha=\sqrt 3$ are those obtained from string theory and Kaluza-Klein compactifications, respectively. Note that $\phi$ and $A_\mu$ here are dimensionless. The dimensionful physical fields are 

\begin{equation}
   \label{physical fields}
       \tilde{\phi}=\sqrt{2}M_P\,\phi; \,\,\,\, \,\,\,\, \tilde{A_{\mu}}=\sqrt{2}M_P\,A_\mu.
   \end{equation}
In the following, for notation simplicity, we will use $\phi$ for both the dimensionful and the dimensionless fields.

A static, spherically symmetric solution to the Einstein's equations was given in \cite{Gao:2004tu} for $V(\phi)$ of the form

\begin{equation}
\label{dilaton potential}
V(\phi)=\frac{2}{3}\frac{\Lambda}{(1+\alpha^2)^2}\left[\alpha^2(3\alpha^2-1)e^{-2\frac{\phi-\phi_0}{\alpha}}+(3-\alpha^2)e^{2\alpha(\phi-\phi_0)}+8\alpha^2e^{\alpha(\phi-\phi_0)-\frac{\phi-\phi_0}{\alpha}}\right],    
\end{equation}
where $\Lambda$ is the cosmological constant and $\phi_0$ the asymptotic value of $\phi(r)$ for $r\to \infty$. 

Contemplating the form of the action in \eqref{Einstein Maxwell dilaton deSitter action}, we can identify the gauge couplings $g =e^{\alpha \phi}$ and its asymptotic value $g_0 =e^{\alpha \phi_0}$. Then \eqref{dilaton potential} can be written as:
\begin{equation}
\label{dilaton potential-g}
V(\phi)=\frac{2}{3}\frac{\Lambda}{(1+\alpha^2)^2}\left[\alpha^2(3\alpha^2-1)\left(\frac{g}{g_0}\right)^{-{2}/{\alpha^2}}+(3-\alpha^2)\left(\frac{g}{g_0}\right)^2+8\alpha^2 \left(\frac{g}{g_0}\right)^{1-{1}/{\alpha^2}}\right].    
\end{equation}
From the sign of the exponent of the coupling in each term, at least for some rational values of $\alpha$, we can associate the first term to non-perturbative contributions, the second to perturbative while the third is non-perturbative for $\alpha <1$, perturbative correction for $\alpha>1$ and a tree-level contribution for $\alpha=1$.  In the latter case, the potential takes the simple form:
\begin{equation}
\label{dilaton potential-g}
V(\phi)=\frac{1}{3} {\Lambda} \left[ \frac{g_0^2}{g^2}+  \frac{g^2}{g_0^2} +4\right].    
\end{equation}
where one could associate the first, second and third terms to  $D$-brane fluxes,  one-loop effect and tree-level cosmological constant contributions, respectively.
One can go further and try to imagine different realizations of such kind of potential in models with flux compactifications. One could start with a gauge theory living on a brane wrapping a cycle of volume $V$ in the internal which has a gauge coupling $g$ that go as $V^{-1/2}$ and assumes that the dilaton $\phi$ measures this volume, $V= e^{- 2 \alpha \phi}$ in string length units. The potential can be written as
\begin{equation}
\label{dilaton potential-g}
V(\phi)=\frac{2}{3}\frac{\Lambda}{(1+\alpha^2)^2}\left[\alpha^2(3\alpha^2-1)\left(\frac{V}{V_{0}}\right)^{1/{\alpha^2}}+(3-\alpha^2)\frac{V_0}{V} +8\alpha^2 \left(\frac{V}{V_{0}}\right)^{({\alpha^2}-1)/{2 \alpha^2}}\right].    
\end{equation}
One can then identify the first and the third terms as resulting form fluxes inside cycles that have smaller or bigger volumes that measure $1/{\alpha^2}$ and $(1-{\alpha^2})/{2 \alpha^2}$ of the volume $V$.

It is straightforward to see that for the (A)dS case the potential has a global (maximum) minimum as long as $1/ \sqrt{3} \le \alpha \le  \sqrt{3}$. It has only a local minimum, a local maximum, and it is unbounded from below for $0< \alpha < 1/\sqrt{3}$ and $\alpha > \sqrt{3}$. For the AdS case, $\Lambda <0$, embeddings of the five-dimensional version of this potential and associated black hole solutions in supergravity, or in string theory, have been briefly discussed in \cite{Elvang:2007ba}. It was pointed out in \cite{Elvang:2007ba} that these correspond to what is known as superstars or giant gravitons, and for some peculiar values of $\alpha$, as $0, 1/\sqrt{3}, 2/\sqrt{3}$ they are consistent solutions of truncated $N=2$ supergravity.

For the dS case, $\Lambda >0$, the situation is more complicate. No asymptotically dS space-time has been constructed from string theory. Present attempts rest mainly on  vacua from supergravity equations of motion with possible presence of non-perturbative contributions from branes. It is not clear if further investigations  including all quantum corrections will allow to construct such solutions. Strictly speaking the conjectures forbid stable solution, thus one could still consider the possibility of long lived vacua for very small values of the cosmological constant as the expected life-time is of order $H^{-1} \log {H}$ (see, for example, \cite{Bedroya:2019snp}). This is an important issue that goes beyond the scope of this paper but should be kept in mind of the reader.


The black hole metric solution of the equations of motion reads 
\begin{equation}
\begin{cases}
\label{metric alpha generic dS}
\mathrm ds^2 =&-\left[\left(1-\frac{r_+}{r}\right)\left(1-\frac{r_-}{r}\right)^{\frac{1-\alpha^2}{1+\alpha^2}}\mp H^2r^2\left(1-\frac{r_-}{r}\right)^{\frac{2\alpha^2}{1+\alpha^2}}\right]\mathrm dt^2  \\
&+\left[\left(1-\frac{r_+}{r}\right)\left(1-\frac{r_-}{r}\right)^{\frac{1-\alpha^2}{1+\alpha^2}}\mp H^2r^2\left(1-\frac{r_-}{r}\right)^{\frac{2\alpha^2}{1+\alpha^2}}\right]^{-1}\mathrm dr^2 \\
&+r^2\left(1-\frac{r_-}{r}\right)^{\frac{2\alpha^2}{1+\alpha^2}}\mathrm d\Omega_2^2,\\
e^{2\alpha\phi}=&e^{2\alpha\phi_0}\left(1-\frac{r_-}{r}\right)^{\frac{2\alpha^2}{1+\alpha^2}},\\
F=& \frac{1}{\sqrt{4\pi G}} \frac{Qe^{2\alpha\phi_0}}{r} \,\mathrm dt\wedge\mathrm dr.
\end{cases}
\end{equation}
Here $H^2$ is the Hubble parameter $H^2={|\Lambda|}/{3}$. When $\Lambda=0$, this reproduces the asymptotically flat black hole solutions of \cite{Gibbons:1987ps,Garfinkle:1990qj}. Otherwise, the solution is either an asymptotically dS (upper sign) or AdS (lower sign) space-time.
The relation between the integration constants $r_+$, $r_-$ and the mass and charge is 

\begin{equation}
\label{relations constants physical parameters}
\begin{cases}
2M=r_+ +\frac{1-\alpha^2}{1+\alpha^2}r_- ,\\
Q^2e^{2\alpha\phi_0}=\frac{r_+ r_-}{1+\alpha^2}, \\
D=\frac{\alpha}{1+\alpha^2}r_-,
\end{cases}
\end{equation}
where $D$ is the scalar charge of the black hole defined as the integral over a two sphere at infinity, $D=\frac{1}{4\pi}{\underset{r\to \infty}{\lim}}\int d^2\Sigma^\mu \nabla_\mu \phi$, or, equivalently, through the expansion $\phi=\phi_0 -\frac{D}{r}+\mathcal O\left(\frac{1}{r^2}\right)$ at large $r$. This family of solutions have only two independent parameters (in addition to the constant asymptotic value $\phi_0$): $r_+,r_-$ or $Q,M$.
Inverting the relations, we obtain $r_+$, $r_-$ from $Q$ and $M$ as 

\begin{equation}
\begin{cases}
\label{inverse relations constants physical parameters}
r_+=M\pm\sqrt{M^2-(1-\alpha^2)Q^2e^{2\alpha \phi_0}} \\
r_-=\frac{(1+\alpha^2)Q^2e^{2\alpha \phi_0}}{M\pm\sqrt{M^2-(1-\alpha^2)Q^2e^{2\alpha \phi_0}}},
\end{cases}    
\end{equation}
and $D$ 

\begin{equation}
D= \alpha\frac{Q^2e^{2\alpha \phi_0}}{M\pm\sqrt{M^2-(1-\alpha^2)Q^2e^{2\alpha \phi_0}}}   
\label{definition scalar charge},
\end{equation}
as long as $(1-\alpha^2)Q^2e^{2\alpha\phi_0} > {M^2}$. In order to map the uncharged $Q=0$ case to the Schwarzschild metric, we choose this limit to correspond exclusively to $r_-=0$.  Contrary to the choice $r_+=0$, this allows the metric \eqref{metric alpha generic dS} to truly take the desired form. Thus, we will consider in the following only solutions with a '$+$' sign in \eqref{inverse relations constants physical parameters} and \eqref{definition scalar charge}.\\

As a consequence of the above relations between $(M,Q)$ and the integration constants $(r_+,r_-)$, some peculiarities arise.
\begin{itemize}
\item When $\alpha\ge1$, probing the $(r_+,r_-)$ plane allows to sweep the entire $(M,Q)$ one. The region $r_+<\left[{(\alpha^2-1)}/{(\alpha^2+1)}\right]r_-$ defines negative masses $M<0$ and is unphysical. A bijection is then defined between the $r_+\ge\left[{(\alpha^2-1)}/{(\alpha^2+1)}\right]r_-$ portion of the $(r_+,r_-)$ plane and the whole $(M,Q)$ one.

\item When $0<\alpha<1$, for $M^2<(1-\alpha^2) Q^2e^{2\alpha\phi_0}$ both the constants $r_+,r_-$ and the metric become complex. A part of the $(M,Q)$ plane is inaccessible to the solution, a manifestation of the fact that

\begin{equation*}
   M^2-(1-\alpha^2)Q^2e^{2\alpha\phi_0}=\left(\frac{r_+}{2}-\frac{1-\alpha^2}{1+\alpha^2}\frac{r_-}{2}\right)^2 
\end{equation*}
 is always positive in the parametric coordinates system $(r_+,r_-)$. There, writing $r_-=r_+\tan\theta$, the charge-to-mass ratio

\begin{equation*}
 \frac{Q^2e^{2\alpha\phi_0}}{M^2}=\frac{4}{1+\alpha^2}\frac{\tan\theta}{\left(1+\frac{1-\alpha^2}{1+\alpha^2}\tan\theta\right)^2}   
\end{equation*}
 monotonically increases from $0$ to ${1}/{(1-\alpha^2)}$ for $\theta\in\left[0,\arctan\frac{1+\alpha^2}{1-\alpha^2}\right]$, reaches its maximal value and then monotonically decreases to 0 for $\theta\in\left[\arctan\frac{1+\alpha^2}{1-\alpha^2},\frac{\pi}{2}\right]$. In this second copy of the $\left(1-\alpha^2\right)Q^2e^{2\alpha\phi_0}<M^2$ parametric region, $Q$ vanishes for $r_+=0$ and, following the discussion below \eqref{definition scalar charge}, we discard it. The bijection is now defined between the $r_+\ge\left[{(1-\alpha^2)}/{(1+\alpha^2)}\right]r_-$ and the $M^2\ge(1-\alpha^2)Q^2e^{2\alpha\phi_0}$ portions of the planes. 
\end{itemize}
Combining them, these observations show that $r_+<\left|{(1-\alpha^2)}/{(1+\alpha^2)}\right|r_-$ defines a non physical region for all $ \alpha \ne 0$. The Reissner-Nordstr\"om solution ($ \alpha = 0$) does not suffer from the same issue, as will be discussed in the next section.






\section{Asymptotically Flat Black Holes: $\Lambda =0$}

For $\alpha=0$,  $g_{00}=-\left(1-\frac{r_+ +r_-}{r}+\frac{r_+r_-}{r^2}\right)$ $=-\left(1-\frac{2M}{r}+\frac{Q^2}{r^2}\right)$, we recover the Reissner-Norsdtr\"om solution. The $r_+$ and $r_-$ constants do not enter separately into the metric anymore but only through the combinations $r_+ +r_-$ and $r_+r_-$. It is thanks to this property that Reissner-Nordstr\"om solutions (either flat or asymptotically (A)dS) do not suffer from the complex valued region discussed above. \\

When $\alpha\ne 0$, $r_-$ is the location of a singular surface while $r_+$ is the only event horizon of the black hole. The condition for the singularity to be shielded by the horizon is simply $r_+>r_-$, that is:
\begin{equation}
\label{dilatonicWGC}
Q^2e^{2\alpha\phi_0} < \left(1+\alpha^2\right)M^2    
\end{equation}  
In this case of asymptotically flat black holes, the complex valued region is beyond the reach of the black hole solution.







\section{Dilatonic de Sitter Black Holes: $\Lambda>0$ }

When $\alpha = 0$, the dilaton decouples and we recover the Reissner-Nordstr\"om-de Sitter solution studied in \cite{Antoniadis:2020xso}. For $\alpha \neq 0$, one needs to distinguish between several cases, corresponding to different behaviours of $g_{00}$. Here, $r_+$ does not determine the location of the horizon anymore, while $r_-$ still indicates the coordinate of a singular surface. \\

\subsection{$\alpha=1$}

The $\alpha=1$ case allows for explicit expressions of the black hole horizons. The metric can be written as 
\begin{equation}
    \label{DdS metric alpha=1}
    \mathrm{ds^2}=-\left(1-\frac{2M}{r}-H^2r(r-2D)\right)\mathrm{d}t^2+\left(1-\frac{2M}{r}-H^2r(r-2D)\right)^{-1}\mathrm{d}r^2+r(r-2D)\mathrm{d}\Omega_2^2.
\end{equation}
 $D$ is related to $M$ and $Q$ through 
\begin{equation}
 D=\frac{Q^2e^{2\phi_0}}{2M}   
\end{equation}
and $r=r_-= 2D$ is a singular surface. The horizons correspond to the loci of the roots of the polynomial of degree 3 in $r$:
\begin{equation}
    P(r)=H^2r^3-2DH^2r^2-r+2M
\end{equation}
Their explicit expression is not very illuminating. We find more instructive, in particular for discussing below $\alpha \neq 1$, to provide a description of the behaviour of the roots as functions of $M$ and $D$. \\

First, note that $P(r) \underset{r\rightarrow+\infty}{\rightarrow} +\infty$, and can have two extrema $R_- < R_+$ given by the roots of $P'(r)=3H^2r^2-4DH^2r-1$. As $R_-  R_+ =-1$, $R_- <0$ while $R_{+}=\frac{2}{3}D+\frac{1}{6}\sqrt{16D^2+\frac{12}{H^2}} >0$. We are interested only in solutions of $P(r)=0$ in the region $r>2D$ outside the singularity. Therefore, we will discuss the signs of $P(2D)$ and, when $R_+ > 2D$, $P(R_+)$.\\

The case of $R_+<2D$, i.e. $D^2H^2>\frac{1}{4}$, corresponds to $r_-^2H^2>1$, which means that the radius of the singular surface is greater than the Hubble's one. No black hole solutions can arise there: when $P(2D)<0$ one root is present, otherwise the polynomial is always positive for all $r>2D$. \\

We restrict from now on to $R_+>2D$. When $P(2D)=2(M-D) \leq 0$, $P$ only has one root. If $P(2D)>0$, there can be 0,1 or 2 roots, depending on the sign of $P$ at the minimum $R_+$.

Studying this case, i.e. $M> D$, we have
\begin{equation}
P(R_+)=-\frac{16}{27}D^3H^2-\frac{2}{3}D+2M-\sqrt{4D^2+\frac{3}{H^2}}\left(\frac{8}{27}D^2H^2+\frac{2}{9}\right).  
\end{equation}
If $-\frac{16}{27}D^3H^2-\frac{2}{3}D+2M$ is negative, the sign of $P(R_+)$ is fixed to be negative, and there are two zeros for $P$. 
In order to further investigate the sign of $P(R_+)$, it is helpful to consider the function

\begin{equation}
\begin{cases}
\label{definition of U(D)}
&U(D)\equiv-\frac{16}{27}D^3H^2-\frac{2}{3}D+2M  \\
&D<D_1 \rightarrow U(D)>0  \\
&D=D_1 \rightarrow U(D)=0  \\
&D>D_1 \rightarrow U(D)<0 
\end{cases}
\end{equation}
We have $U(0)=2M>0$ and $U$ is decreasing with $D$. There is so one solution $D_1$ such that $U(D_1)=0$. For $D>D_1$, $U$ is negative and $P$ has two zeros. The region $D<D_1$, where $U(D)$ is positive, needs further investigation. 

It is easier for the rest of the computation, with $D<D_1$, to reformulate the zeros of $P(R_+)$ as the zeros of a simpler function: 
\begin{align*}
P(R_+)=0 &\Leftrightarrow    \left(-\frac{16}{27}D^3H^2-\frac{2}{3}D+2M\right)^2=\left(4D^2+\frac{3}{H^2}\right)\left(\frac{8}{27}D^2H^2+\frac{2}{9}\right)^2\\
&\Leftrightarrow -\left(\frac{4}{3}\right)^3H^2MQ(D)=0,
\end{align*}
where $Q$ is a function of $D$ defined as 

\begin{equation}
\label{definition of Q(D)}
Q(D)=D^3+\frac{1}{16H^2M}D^2+\frac{9}{8H^2}D-\frac{27M}{16H^2}+\frac{1}{16H^4M},   
\end{equation}
so that $P(R_+)<0$ when $Q(D)>0$. $Q$ is an increasing function for  positive $D$. The sign of $Q(0)=-\frac{27M}{16H^2}+\frac{1}{16H^4M}$ discriminates between two cases. If $Q(0)>0$, $Q(D)$ is positive for all positive $D$. If $Q(0)<0$, there is one $D_0$ such that $Q(D_0)=0$:
\begin{equation}
\begin{cases}
&D<D_0 \rightarrow Q(D)<0 \Rightarrow P(R_+)>0 \Rightarrow P(r)\neq 0, \, \,  \forall r   \in {\mathbb R^+} \nonumber \\
&D=D_0 \rightarrow Q(D)=0 \nonumber \\
&D>D_0 \rightarrow Q(D)>0 \Rightarrow P(R_+)< 0.
\end{cases}
\end{equation}

Imposing the necessary condition $D^2H^2<\frac{1}{4}$, we shall now group  all cases. There are three possibilities corresponding to 0, 1 or 2 roots.\\
\begin{itemize}
    \item For $P$ to have 2 roots, the first condition to be satisfied is $P(2D) > 0$, i.e. $D < M$. If $D>D_1$, $P$ has two roots and there is no need for further investigations. If $D<D_1$, $P$ is also assured to have two roots when $M^2H^2<\frac{1}{27}$. On the contrary, when $M^2H^2\ge\frac{1}{27}$, $P$ has two roots when the additional condition $Q(D)>0\Leftrightarrow D>D_0$ is met.\\

    \item There are two scenarios where $P$ has one root. The first is realized when $P(2D) \leq 0$, corresponding to $D\geq M$. As $P(0)>0$, this happens when one of the two roots above is behind the singularity. The second scenario is met when $P(2D)\ge0$ and $P(R_+)=0$, corresponding to $M\ge D$ and $D=D_0$ with $D<D_1$. This latter case is found when the two horizons discussed in the previous point coincide.\\

    \item Finally, the case where $P$ does not have roots corresponds to $D<M$, $D<D_1$, $D<D_0$ and $M^2H^2>\frac{1}{27}$.\\

\end{itemize}
All the cases listed above depend on the values of $D_0$ and $D_1$. Those are given in terms of $M$ as roots of the polynomials $Q$ and $U$. More compact expressions, that we present below, can be given using the variables $Y=DH$ and $X=MH$. This gives: 
\begin{align}
Y_0=&-\frac{1}{48X}+\frac{1}{48}\left(-\frac{1}{X^3}-\frac{2^4.3^3.5}{X}+2^7.3^6X+\frac{48\sqrt{3}}{X^2}\sqrt{1+2^2.3^4X^2+2^4.3^7X^4+2^6.3^9X^6}\right)^{\frac{1}{3}} \nonumber \\
&-\frac{16}{3}\! \!  \left(\frac{27}{8}-\! \!  \frac{1}{256X^2} \! \! \right)\! \!  \left(-\frac{1}{X^3}-\frac{2^4.3^3.5}{X}+2^7.3^6X+\frac{48\sqrt{3}}{X^2}\sqrt{1+2^2.3^4X^2+2^4.3^7X^4+2^6.3^9X^6}\right)^{-\frac{1}{3}} \nonumber \\
Y_1=&\frac{3}{2\sqrt{6}}\left(-3\sqrt{6}X+\sqrt{1+54X^2}\right)^{-\frac{1}{3}}-\frac{3}{2\sqrt{6}}\left(-3\sqrt{6}X+\sqrt{1+54X^2}\right)^{\frac{1}{3}}
\end{align}
Actually, the value of $Y_0$ presented just above is complex for $X<\frac{1}{12\sqrt{6}}$. In that range of parameters, of the three roots of $Q(D)$, it is another one which is real, corresponding to $Y_0$ with an absolute value taken on the factors elevated to the $\pm\frac{1}{3}$ power and on the factor $\frac{27}{8}-\frac{1}{256X^2}$. However, as we are only interested in $D>0$ and $D_0$ is positive only for $X>\frac{1}{\sqrt{27}}$, the expression for $Y_0$ given above is real in the whole range of interest for $D$ and the absolute values are of no use. \\

\begin{figure}
    \centering
    \includegraphics[scale=0.7]{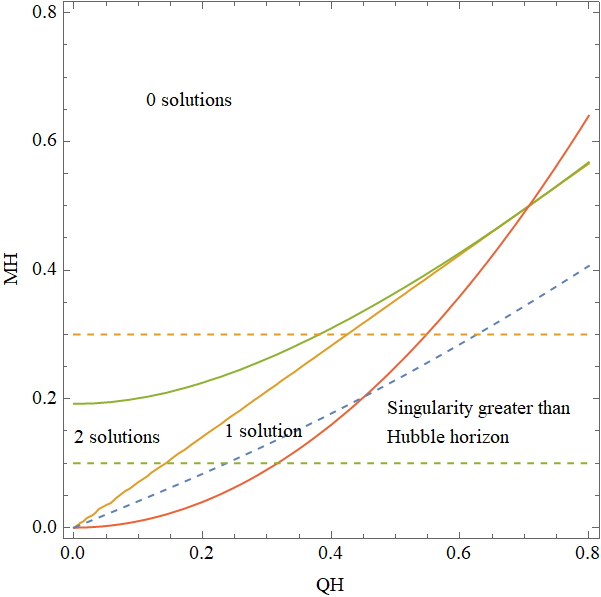}
    \caption{Number of horizons of the $\alpha =1$ de Sitter black hole as a function of $M H$ and $Q H$. The green curve represents $H D_0$, the yellow one the limit $\sqrt{2}MH=Qe^{\alpha\phi_0}H$, and the red one $D^2H^2=\frac{1}{4}$. Dotted lines are for intermediate steps and discussions in the text.}
    \label{fig:my_label}
\end{figure}

The different cases for the black hole horizons are represented graphically in figure $\ref{fig:my_label}$. Instead of $D$, we used the electric charge $Q$ (actually, $Q$ really is $Qe^{\phi_0}$) to define $x$-axis.
The green curve represents $D=D_0$, while the yellow one is $M=D$. The function $D_1$, represented by the dashed blue curve, is below $M=D$ for $D^2H^2<\frac{1}{4}$ so that, according to our previous findings, it plays no role in the separation of the different regimes. In the region between the green and the yellow curves, black hole solutions with two horizons are found. For $Q=0$, the discriminant between solutions with two and zero horizons is $M=\frac{1}{\sqrt{27}H}$, as it should. 
Solutions describing a naked singularity with a cosmological horizon are found below the yellow curve. Finally, the red curve is defined by $2DH=1$. On its right, the radius of the singularity is greater than the Hubble's. \\

To illustrate the solution, we now follow two straight horizontal lines, like the green and yellow dashed ones, in figure \ref{fig:my_label}, with $MH<\frac{1}{\sqrt{27}}$ in one case and $\frac{1}{\sqrt{27}}<MH\le\frac{1}{2}$ in the other.  
\begin{itemize}
    \item $\boldsymbol{MH<\frac{1}{\sqrt{27}}}$. At $Q=0$ there are two horizons: the event and the cosmological horizon. As $Q$ grows, the radius of the cosmological horizon and of the singularity increase while that of the event horizon decreases until the value $Qe^{\phi_0}=\sqrt{2}M$ is reached. Here, the event horizon coincides with the singularity. For $Qe^{\phi_0}>\sqrt{2}M$, only the cosmological horizon surrounds $r_-=\frac{Q^2e^{2\alpha\phi_0}}{M}$. The radius of the singular surface increases with $Q$ until it meets the Hubble radius when $Q^2e^{2\alpha\phi_0}=\frac{M}{H}$.

    \item $\boldsymbol{\frac{1}{\sqrt{27}}<MH\le\frac{1}{2}}$. With $MH<\frac{1}{2}$, at $Q=0$ no horizons are present. This remains true until the condition $\frac{Q^2e^{2\alpha\phi_0}}{2M}=D_0$ is met: at this point, one horizon appears. Here, two roots of $g_{00}$ coincide, meaning that the event and cosmological horizons have the same size. As $Q$ further grows the two horizons disentangle, the radius of the event horizon shrinks, while that of the cosmological horizon increases. From now on, the analysis is the same as in the previous point: when the condition $Qe^{\alpha\phi_0}=\sqrt{2}M$ is reached, the singularity merge with the event horizon, and for greater charges the solutions only show a cosmological horizon.\\ When $MH=\frac{1}{2}$, the region with two horizons disappears. At the point $MH=\frac{1}{2}$, $Qe^{\alpha\phi_0}H=\sqrt{2}MH=\frac{1}{\sqrt 2}$, the green, yellow and red curves meet. Here, both the two roots of $g_{00}$ coincide with the singularity that coincides, in turn, with the Hubble horizon. Put simply, the locations of the singularity, the event, the cosmological and the Hubble horizons all coincide. In terms of the previously defined quantities, this corresponds to the case where $P(2D)=P(R_+)=0$ with $R_+=2D$. This point defines the maximal mass and charge for which a black hole solution exists. Larger charges allow for the presence of a cosmological horizon, with the singularity bigger than the Hubble surface.  

\end{itemize}

For $MH>\frac{1}{2}$, no black hole solution is possible: the singularity is either naked, when $Qe^{\alpha\phi_0}<\sqrt{2}M$, or shielded by a cosmological horizon when $Qe^{\alpha\phi_0}>\sqrt{2}M$, with the latter coinciding with the singularity when the equality is verified. The condition for the singularity to be bigger than the Hubble horizon is now met before this last one. \\

If we follow the arguments of \cite{Antoniadis:2020xso} to infer the WGC condition from the absence of event horizons shielding the singularity, the WGC would require the existence of a state with mass $m$ and charge $q$, in geometrized units, satisfying $qe^{\phi_0}>\sqrt{2}m$. This corresponds to the dilatonic WGC bound in asymptotically flat space-time, as discussed above. Thus, for $\alpha =1$, the dilatonic WGC seems to be insensitive to the presence of a cosmological constant.

\subsection{$\alpha>1$}

We first study the $\alpha\to \infty$ limit, where one should recover the Schwarzschild-de Sitter solution, and then look at the generic $\alpha>1$ case.\\

In the $\alpha \to \infty$ limit the metric reads

\begin{equation}
\mathrm ds^2=-\left[\frac{1-\frac{r_+}{r}}{1-\frac{r_-}{r}}-H^2r^2\left(1-\frac{r_-}{r}\right)^2\right]\mathrm dt^2 +\left[\frac{1-\frac{r_+}{r}}{1-\frac{r_-}{r}}-H^2r^2\left(1-\frac{r_-}{r}\right)^2\right]^{-1}\mathrm dr^2 +r^2\left(1-\frac{r_-}{r}\right)^2\mathrm d\Omega_2^2.   
\end{equation}
To study the horizons of the above metric, we need to find the roots of the polynomial 

\begin{equation}
G(r)\equiv H^2(r-r_-)^3-(r-r_+)=0.   
\end{equation}
$G$ has two extrema: a minimum at $r=r_-+\frac{1}{\sqrt 3 H}$ and a maximum at $r=r_--\frac{1}{\sqrt 3 H}$. The latter is inside the singular surface. The knowledge of the values on the singular surface, $G(r_-)$, and at its minimum, $G(r_-+\frac{1}{\sqrt 3 H})$, allows to find the number of roots of $G$. We have

\begin{equation}
\begin{cases}
G(r_-)=r_+-r_- \\
G(r_-+\frac{1}{\sqrt 3 H})=r_+-r_- -\frac{2}{\sqrt{27}H}.
\end{cases}
\end{equation}
For $0< r_+ -r_-<\frac{2}{\sqrt{27}H}$, the singularity is protected by two horizons: the event and the cosmological horizons. When $r_+-r_-=\frac{2}{\sqrt{27} H}$, the two horizons merge. Above, neither the event nor the cosmological horizon are present. At $r_+=r_-$, the event horizon coincides with the singularity. Using \eqref{inverse relations constants physical parameters} one obtains, for $\alpha\to\infty$, $r_+-r_-=2M$. Thus, we discard the $r_+<r_-$ region as corresponding to negative masses. In the $\alpha\to \infty$ limit of \eqref{metric alpha generic dS} the Schwarzschild-de Sitter solution is thus recovered: the discriminant between a naked and a shielded singularity is the sign of $M-\frac{1}{\sqrt{27} H}$. 
\\

Now, we consider the general metric \eqref{metric alpha generic dS} and define a new function $F$ that vanishes for the same values of $r$ than $g_{00}$: 

\begin{equation}
\label{polynome for alpha2>1/3}
F(r)\equiv r-r_+-H^2r^3\left(1-\frac{r_-}{r}\right)^{\frac{3\alpha^2-1}{\alpha^2+1}}.    
\end{equation}

To investigate the solutions of $F(r)=0$, we divide $F$ into the sum of two contributions: $A(r)\equiv r-r_+$, and $B(r)\equiv H^2r^3\left(1-\frac{r_-}{r}\right)^{\frac{3\alpha^2-1}{\alpha^2+1}}$. \\
The intersection points of the two curves defined by $A$ and $B$ give the zeros of $F$. We carry the analysis in two regions of the parameter space:

\begin{itemize}
    \item For $r_+ \leq r_-$, $A(r_-) \geq 0$ and the two curves always cross in one point. Accordingly there is, in this case, only one zero, corresponding to the cosmological horizon.
    
    \item For $r_+>r_-$, $A(r_-)<0$ and there are either two, one or zero solutions depending on the location of the point $r_0$ where $B'(r_0)=1$. 
    $B(r_0)\le A(r_0)$ corresponds to the case where the function has two zeros, coalescing into one when the equality is satisfied. $B(r_0)> A(r_0)$ will determine the horizon-less regime where the dS space-time causal patch  has been completely eaten by the black hole.
\end{itemize}
We consider, from now on, $r_+ > r_-$. To discriminate between the different regimes we just described, we proceed in the following way.\\

First, we observe that the limit for the two zeros to collapse into one is obtained where $A(r)$ and $B(r)$ are tangent, thus $F(r_0)=0$ and $F'(r_0)=0 \left(B'(r_0)=1\right)$. Consider\footnote{The solutions of the system are always two as the equation $F(r)=0$, with the prior $F'(r)=0$, reduces to a quadratic equation for $r$.} $r_{0\,\pm}$ two functions of $r_{\pm}$ given by:

\begin{equation}
\label{rzeroplus and rzerominus}
r_{0\,\pm}=\frac{(3-\alpha^2)r_-+3(1+\alpha^2)r_+}{4(1+\alpha^2)}\pm\sqrt{\left(\frac{(3-\alpha^2)r_-+3(1+\alpha^2)r_+}{4(1+\alpha^2)}\right)^2-2\frac{r_+r_-}{1+\alpha^2}}.    
\end{equation}
When 
\begin{equation}
\begin{cases}
\label{strategy to compute the bound on BHs}
F(r_{0\pm})=0  \\
F'(r_{0\pm})=0
\end{cases}
\end{equation}
the event and cosmological horizons coincide. As $F(r) \rightarrow -\infty$ for $r \rightarrow \infty$, starting with $F(r_-)<0$, if $F(r)$ takes a positive value at some coordinate value this ensures that it crosses twice the abscissa axis thus allowing the existence of two horizons. Therefore, when $r_{0\pm}$ are both greater than $r_-$, $r_+ > r_-$, the black hole solution exists in the parameter region of $(r_+, r_-)$ where $F(r_{0+})>0$ and $F(r_{0-})<0$. \\

Next, note that for $\alpha>1$, $r_{0-}<r_-$ does not intervene. Using $F(r_-)<0$, the region where two horizons are present is defined by $F(r_{0+})>0$.  
In terms of $M$ and $Q$, $F(r_{0+})>0$ translates to 
\begin{align}
    &\left( (1-2\alpha^2)M+(\alpha^2-2)\sqrt{M^2-(1-\alpha^2)Q^2e^{2\alpha\phi_0}}+\sqrt{P(M,Q,\alpha,\phi_0)}\right)\nonumber\\
    &-H^2\left((10+9\alpha^2)M-(4+3\alpha^2)\sqrt{M^2-(1-\alpha^2)Q^2e^{2\alpha\phi_0}}+\sqrt{P(M,Q,\alpha,\phi_0)} \right)^{\frac{3\alpha^2-1}{1+\alpha^2}}\times\nonumber \\
    &\left((4+3\alpha^2)M-\sqrt{M^2-(1-\alpha^2)Q^2e^{2\alpha\phi_0}}+\sqrt{P(M,Q,\alpha,\phi_0)} \right)^{\frac{4}{1+\alpha^2}} >0,
\end{align}
where $P(M,Q,\alpha,\phi_0)$ is defined by 
\begin{equation}
 P(M,Q,\alpha,\phi_0)=(17+24\alpha^2+9\alpha^4)M^2-(9+15\alpha^2+8\alpha^4)Q^2e^{2\alpha\phi_0}-(8+6\alpha^2)M\sqrt{M^2-(1-\alpha^2)Q^2e^{2\alpha\phi_0}}.   
\end{equation}
\begin{figure}
    \centering
    \includegraphics[scale=0.7]{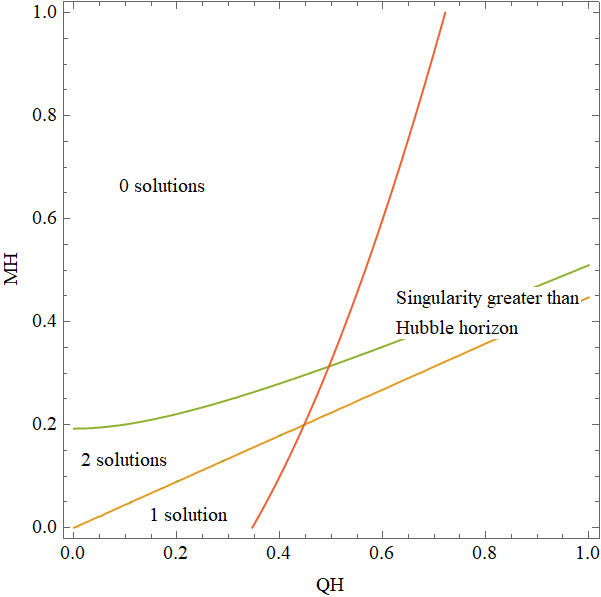}
    \caption{Number of horizons of the $\alpha >1$ de Sitter black hole as a function of $M H$ and $Q H$. The green curve represents $F(r_{0+})=0$, the yellow one the limit $Q^2=(1+\alpha^2)M^2$, and the red one $r_- H=1$.  We chose for the illustration $\alpha =2$ and $\phi_0=0$.}
    \label{graphe alpha grand}
\end{figure}
Figure \ref{graphe alpha grand} presents the results of the discussion above. In analogy to the case $\alpha=1$, we have added a further constraint for the singularity to be inside the Hubble horizon, $r_-<\frac{1}{H}$. 
When the black hole charge vanishes, the equation $F(r_{0+})=0$ reduces to $M=\frac{1}{\sqrt{27}H}$. Consider in this figure a point in the region corresponding to a black hole with two horizons and vary the charge or the mass:

\begin{itemize}
    \item Increasing the mass, the event horizon reaches the cosmological one for the black hole mass $M$ such that $F(r_{0+})=0$. Beyond this value, the singularity is naked.

    \item Increasing the charge, instead, we encounter at some point the line $Q^2e^{2\alpha\phi_0}=(1+\alpha^2)M^2$, where the event horizon and the singular surface $r_-$ coalesce. Continuing to increase the charge, the electromagnetic energy density becomes  strong enough to prevent the formation of an event horizon. The WGC states are expected to have mass and charge in this region of parameters.
\end{itemize}  

  The WGC would then require the existence of a state with $q^2e^{2\alpha\phi_0}>(1+\alpha^2)m^2$, as in the case $\alpha=1$. Although the presence of a cosmological constant changes the form of $g_{00}(r)$, in the $\alpha>1$ case the weak gravity bound would take the same form as in asymptotically flat space-time.

\subsection{$\alpha<1$}

For $\alpha<1$, both the terms in $g_{00}$ given by: 
\begin{equation*}
g_{00}(r)=-\left[\left(1-\frac{r_+}{r}\right)\left(1-\frac{r_-}{r}\right)^{\frac{1-\alpha^2}{1+\alpha^2}}-H^2r^2\left(1-\frac{r_-}{r}\right)^{\frac{2\alpha^2}{1+\alpha^2}}\right]   
\end{equation*}
vanish when $r\to r_-$. For $\alpha^2=\frac{1}{3}$, the $r_-$ dependence factorizes and $g_{00}$ can be written as 

\begin{equation}
\label{g00 alpha2=1/3}
g_{00}(r)\Big|_{\alpha^2=\frac{1}{3}}=-\left(1-\frac{r_-}{r}\right)^{\frac{1}{2}}\left(1-\frac{r_+}{r}-H^2r^2\right),
\end{equation}
where the second factor takes the form of the $g_{00}$ of a Schwarzschild-de Sitter metric with mass $M^*\equiv\frac{r_+}{2}$. The relative importance of the two terms in $g_{00}$ depends on whether $\alpha^2<\frac{1}{3}$ or $\alpha^2>\frac{1}{3}$. A priori, we may expect a dilatonic-like black hole behaviour for $\frac{1}{\sqrt 3}<\alpha<1$, similar to the $\alpha>1$,  while a different, de Sitter-like black hole, behaviour for $\alpha<\frac{1}{\sqrt 3}$. We thus split the $\alpha<1$ analysis in three parts: $\frac{1}{\sqrt 3}<\alpha<1$, $\alpha=\frac{1}{\sqrt 3}$ and $\alpha<\frac{1}{\sqrt 3}$. 

\subsubsection{$\frac{1}{\sqrt 3}<\alpha<1$}

Factorizing the $\left(1-\frac{r_-}{r}\right)$ term, we study the zeros of $F$ defined in \eqref{polynome for alpha2>1/3}, with $3\alpha^2-1>0$. The only difference with the $\alpha>1$ case comes from the convergence of the first term in $g_{00}$ when $r\to r_-$. The convergence is to $0^+$ when $r_+>r_-$ and to $0^{-}$ when $r_+<r_-$. Concretely, this does not affect the zeros of $F$, and thus the results obtained in the case $\alpha \geq 1$. In the $(Q,M)$ plane, the boundaries of the region allowing black holes is still given by $Q^2e^{2\alpha\phi_0}=(1+\alpha^2)M^2$ and $F(r_{0+})=0$. Note that the most involved part of the analysis in the case $\alpha >1$ was for the situation $r_+>r_-$ and used $r_{0-}<r_-$. While
 in the region $\alpha \in\left]\frac{1}{\sqrt 3},1\right[$, $r_{0-}$ can become greater than $r_-$, this only happens when $r_+<r_-$ and therefore does not modify that analysis. 
Black hole arguments would again indicate for the WGC the existence of a particle satisfying $q^2e^{2\alpha\phi_0}>(1+\alpha^2)m^2$. As long as the second term in $g_{00}$ (\textit{de Sitter-like}) is sub-dominant, the transition between black holes and naked singularities (with a cosmological horizon) seems to happen in the same parametric region as in asymptotically flat space-time.

\subsubsection{$\alpha=\frac{1}{\sqrt 3}$}
The $\alpha=\frac{1}{\sqrt{3}}$ case allows for explicit expressions of the horizons and can be studied in full details.  The second factor in \eqref{g00 alpha2=1/3} can be seen as the time component of a Schwarzschild-de Sitter metric with an effective mass $M^*\equiv r_+/2$. This factor has two zeros for $r_+<\frac{2}{\sqrt{27} H}$, degenerate for $r_+=\frac{2}{\sqrt{27} H}$, and none for $r_+>\frac{2}{\sqrt{27} H}$. The roots of the polynomial $P(r)\equiv r-r_+-H^2r^3$ are:

\begin{equation}
\label{horizonsalpha1/3}
\begin{cases}
r_c=\frac{1}{H}\left(\frac{\left(\frac{2}{3}\right)^{1/3}}{\left(-9r_+H+\sqrt 3\sqrt{-4+27r_+^2H^2}\right)^{1/3}}+\frac{\left(-9r_+H+\sqrt 3\sqrt{-4+27r_+^2H^2}\right)^{1/3}}{2^{1/3}3^{2/3}}\right) \\
r_h=-\frac{1}{H}\left(\frac{\left(\frac{2}{3}\right)^{1/3}e^{-i\pi/3}}{\left(-9r_+H+\sqrt 3\sqrt{-4+27r_+^2H^2}\right)^{1/3}}+\frac{\left(-9r_+H+\sqrt 3\sqrt{-4+27r_+^2H^2}\right)^{1/3}e^{i\pi/3}}{2^{1/3}3^{2/3}}\right) \\
r_{--}=-\frac{1}{H}\left(\frac{\left(\frac{2}{3}\right)^{1/3}e^{i\pi/3}}{\left(-9r_+H+\sqrt 3\sqrt{-4+27r_+^2H^2}\right)^{1/3}}+\frac{\left(-9r_+H+\sqrt 3\sqrt{-4+27r_+^2H^2}\right)^{1/3}e^{-i\pi/3}}{2^{1/3}3^{2/3}}\right),
\end{cases}
\end{equation}
where $r_c$, $r_h$ are the cosmological and the event horizons, respectively, and $r_{--}$ is negative thus of no physical interest. In fact, one can see from the coefficients of $P$ that the product of the roots is $-\frac{r_+}{H^2}<0$ and their sum is null. So, there are either two real positive roots and a negative one (corresponding to the case where we have two horizons) or two complex conjugate and a negative root (corresponding to the case where no horizon is present). The transition between these two regimes happens when the horizons coincide, $r_c=r_h$, i.e. for $r_+=\frac{2}{\sqrt{27}H}$. We also require that these roots are located outside the singular surface at $r_-$. In order to study the behaviour of the roots of $g_{00}$, we consider the equations: 
\begin{equation}
 \begin{cases}
 \label{region graph alpha 1/3}
 r_c-r_-=0\\
 r_h-r_-=0\\
 r_+=\frac{2}{\sqrt{27}H}.
 \end{cases}   
\end{equation}

\begin{figure}[t]
    \centering
    \includegraphics[scale=0.6]{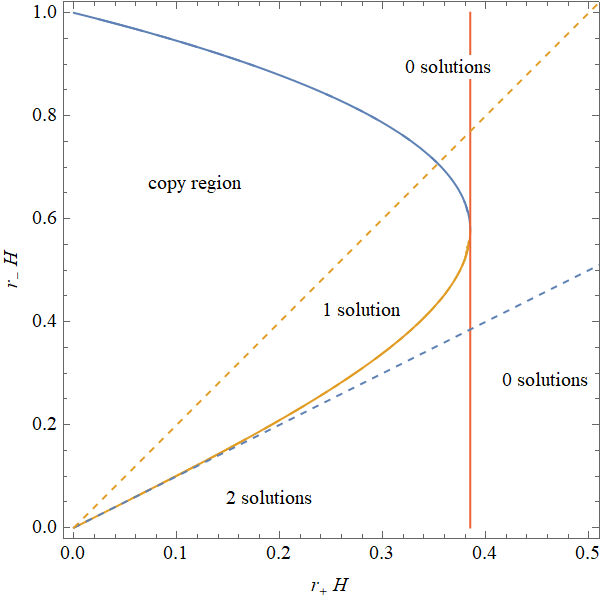}
    \includegraphics[scale=0.6]{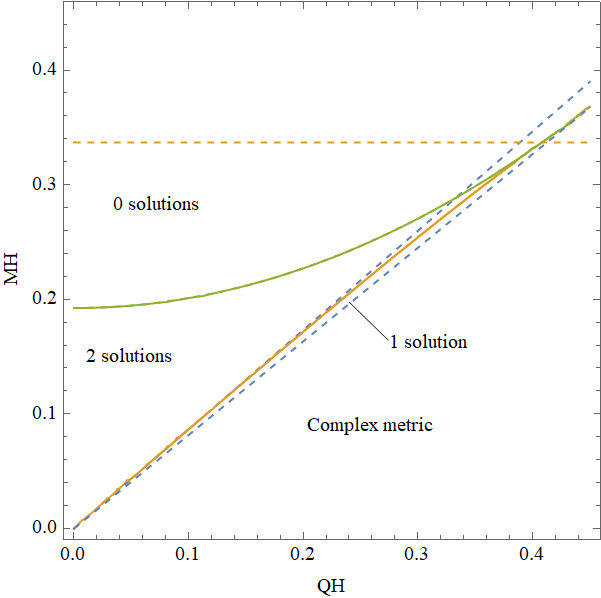}
    \caption{Number of horizons for $\alpha^2=\frac{1}{3}$ in $(r_+H,r_-H)$ (\textit{left}) and $(Q H, M H)$ coordinates (\textit{right}). \\ \textit{Left}: The region below the blue curve corresponds to $r_c>r_-$ and the region below the yellow one to $r_h>r_-$. The intersection of the two regions is populated by black hole solutions. On its right, the dS space-time causal patch is completely eaten. The blue dotted curve is $r_+=r_-$, the yellow dotted one is $r_+={r_-}/{2}$. The portion of plane above the latter does not have counterpart in $(Q H, M H)$. \textit{Right}: Green and yellow curves translate, respectively, the blue and yellow ones from left panel. Blue dotted curves are
    $Q^2=({4}/{3})M^2$ and $Q^2=({3}/{2})M^2$. The union of the region between the yellow curve and the lower dotted blue one from $(0,0)$ to $\left(\frac{1}{\sqrt 6},\frac{7}{12\sqrt{3}}\right)$ and the one between the green and the same dotted line from $\left(\frac{1}{\sqrt 6},\frac{7}{12\sqrt{3}}\right)$ to $\left(\frac{\sqrt 3}{4},\frac{1}{2\sqrt 2}\right)$ give the region with a singularity surrounded only by a cosmological horizon. The yellow dotted line represents the maximal black hole mass: $M_{\mathrm{max}}=\frac{7}{12\sqrt{3}H}$.}
    \label{figure dS alpha^2 un tiers}
\end{figure}

We can see in figure \ref{figure dS alpha^2 un tiers} (\textit{left} panel) the different regimes for $g_{00}$ in the $(r_+H,r_-H)$ plane. It is instructive to understand the $(r_+H,r_-H)$ diagram before moving to the physical parameters $M$ and $Q$. The red curve is $r_+=\frac{2}{\sqrt{27}H}$, while the blue and yellow ones represent $r_c-r_-$ and $r_h-r_-$, respectively. They intersect each other at the point $(r_+,r_-)=\left(\frac{2}{\sqrt{27}H},\frac{1}{\sqrt 3 H}\right)$. 
As long as $r_+<\frac{2}{\sqrt{27}H}$, $r_c>r_-$ is realized below the blue curve and $r_h>r_-$ below the yellow one. As a consistency check of the method used for generic $\alpha$, one can verify here, thanks to the explicit expressions of the horizons, the equivalence of the condition $r_c\ge r_-,\, r_+\le\frac{2}{\sqrt{27}H}$ with $F(r_{0+})\ge0$ and of $r_h\ge r_-,\, r_+\le\frac{2}{\sqrt{27}H}$ with $F(r_{0-})\le 0$. The different regimes are related in the following way: 

\begin{itemize}
    \item The transition between a black hole solution and  a naked singularity with no cosmological horizon happens for $r_+=\frac{2}{\sqrt{27}H}$ when $r_-\le \frac{1}{\sqrt3 H}$.
    
    \item  A transition from the black hole to a naked singularity with a cosmological horizon can only happen in the combined interval $r_+\in\left[0,\frac{2}{\sqrt{27}H}\right]$, $r_-\in\left[0,\frac{1}{\sqrt{3}H}\right]$ when we cross the yellow curve representing $r_h=r_-$. We find here a new bound compared to the $r_+=r_-$ (blue dashed one) present for $\alpha^2>\frac{1}{3}$.
    
    \item The regions defining a naked singularity with or without a cosmological horizon meet on the blue curve when $r_->\frac{1}{\sqrt 3 H}$.
\end{itemize}
The region of existence of the black hole is larger than what it would have been if it was bounded by $r_+=r_-$ $\left(Q^2e^{2\alpha\phi_0}=\frac{4}{3}M^2\left(=(1+\alpha^2)M^2\right)\right)$. The line $r_+={r_-}/{2}$, corresponding to $Q^2=({3}/{2})M^2$ $\left(={M^2}/{(1-\alpha^2)}\right)$, is also shown in the diagram (yellow dashed one). It entirely lies in a region where the singularity is naked. Along this dotted line, the separation between the regions where the singularity is surrounded or not by a cosmological horizon is given by its intersection with the blue curve. The problematic region of a complex valued metric limiting the definition of the coordinates does not intervene in the black hole region. It can only be reached after the singularity has been exposed. Note that each value of $(QH,MH)$ corresponds to two choices of the coordinates $(r_+,r_-)$, one of which is above the line $r_+={r_-}/{2}$ and one below it. However, only this last region leads to black hole solutions.\\

A similar analysis can be carried in terms of the mass and charge parameters $(M,Q)$, using the equations $r_+=M+\sqrt{M^2-\frac{2}{3}Q^2e^{2\alpha\phi_0}}$, $r_-=\frac{4}{3}{Q^2}/{\left(M+\sqrt{M^2-\frac{2}{3}Q^2e^{2\alpha\phi_0}}\right)}$. It is convenient to use $\hat M\equiv MH$ and $\hat Q\equiv e^{\alpha\phi_0}QH$, and the conditions \eqref{region graph alpha 1/3} then take the following form

\begin{align}
\label{cosmological horizon greater than singularity}
r_c-r_-=0 \Leftrightarrow&  \frac{\left(\frac{2}{3}\right)^{1/3}}{\left(-9\left(\hat M+\sqrt{\hat M^2-\frac{2}{3}\hat Q^2}\right)+\sqrt 3\sqrt{-4+27\left(\hat M+\sqrt{\hat M^2-\frac{2}{3}\hat Q^2}\right)}\right)^{1/3}}   \nonumber\\
&+\frac{\left(-9\left(\hat M+\sqrt{\hat M^2-\frac{2}{3}\hat Q^2}\right)+\sqrt 3\sqrt{-4+27\left(\hat M+\sqrt{\hat M^2-\frac{2}{3}\hat Q^2}\right)}\right)^{1/3}}{2^{1/3}\,3^{2/3}} \nonumber \\
&=\frac{4\hat Q^2}{3\left(\hat M+\sqrt{\hat M^2-\frac{2}{3}\hat Q^2}\right)},
\end{align}

\begin{align}
\label{horizon greater than singularity alpha2=1/3}
r_h-r_-=0 \Leftrightarrow&  \frac{\left(\frac{2}{3}\right)^{1/3}e^{-i\pi/3}}{\left(-9\left(\hat M+\sqrt{\hat M^2-\frac{2}{3}\hat Q^2}\right)+\sqrt 3\sqrt{-4+27\left(\hat M+\sqrt{\hat M^2-\frac{2}{3}\hat Q^2}\right)}\right)^{1/3}}   \nonumber\\
&+\frac{e^{i\pi/3}\left(-9\left(\hat M+\sqrt{\hat M^2-\frac{2}{3}\hat Q^2}\right)+\sqrt 3\sqrt{-4+27\left(\hat M+\sqrt{\hat M^2-\frac{2}{3}\hat Q^2}\right)}\right)^{1/3}}{2^{1/3}\,3^{2/3}} \nonumber \\
&=-\frac{4\hat Q^2}{3\left(\hat M+\sqrt{\hat M^2-\frac{2}{3}\hat Q^2}\right)},
\end{align}
and 

\begin{equation}
r_+=\frac{2}{\sqrt{27}H} \Leftrightarrow \hat M = \frac{1}{\sqrt{27}}+\frac{\sqrt 3\hat Q^2}{2}.     
\end{equation}
It is possible to identify the \textit{triple point} $(r_+,r_-)=\left(\frac{2}{\sqrt{27}H},\frac{1}{\sqrt 3 H}\right)$ where the regions with two, one and zero solutions meet with $(\hat Q, \hat M)=\left(\frac{1}{\sqrt 6}, \frac{7}{12\sqrt{3}}\right)$.
We now have all the elements to understand the phase diagram in the $(\hat Q, \hat M)$ plane as displayed in Figure \ref{figure dS alpha^2 un tiers} (\textit{right} panel):

\begin{itemize}
    \item Restricting to masses below the {\it triple point}, thus $\hat M <\frac{7}{12\sqrt{3}}$, the upper bound on the mass allowing the two-horizons solution is given by $r_+=\frac{2}{\sqrt{27}H}$, as shown in the $(r_+H, r_-H)$ plane. It is represented by the green curve, $\hat M=\frac{1}{\sqrt{27}}+\frac{\sqrt 3\hat Q^2}{2}$, from $\left(0,\frac{1}{\sqrt{27}}\right)$ up to the triple point. Above it, the bound from \eqref{cosmological horizon greater than singularity} is stronger and the green curve now corresponds to $r_c=r_-$ .
    
    \item The lower bound on this black hole region is given by \eqref{horizon greater than singularity alpha2=1/3} and it is  represented in yellow in figure \ref{figure dS alpha^2 un tiers}. It corresponds to the limit where the event horizon coincides with the singularity. Note that the lower bound yellow curve starts at $(0,0)$ and crosses the upper bound green curve at the \textit{triple point}. After that, the yellow curve runs above the green one and does not bound any physical region. 
\end{itemize}
The graphical representation of the two bounds reveals that for masses $M>\frac{7}{12\sqrt 3 H}$ the event horizon cannot form: this is the maximal mass above which asymptotically de Sitter black hole solutions are no more possible (yellow dashed line). Accordingly, this point corresponds to a maximal charge $Q_{\mathrm{max}}=\frac{1}{\sqrt{6}H}$. 
\begin{itemize}
  \item  Singularities with only a cosmological horizon are found in the domain given by the union of: (1) the region between the yellow and the lower dashed blue curve from the origin up to the \textit{triple point}, (2) the region between the green and the same dashed blue curve, now above it. On this dashed curve $Q^2e^{2\alpha\phi_0}=({3}/{2})M^2$ marks the limit of definition of the metric.
\end{itemize}  

The  ($r_c=r_-$) green curve crosses the blue one, corresponding to $Q^2e^{2\alpha\phi_0}=({3}/{2})M^2$ in the point $\left(Qe^{\alpha\phi_0},M\right)=\left(\frac{ \sqrt 3}{4 H},\frac{1}{2\sqrt{2}H}\right)$. This is a point of maximal charge and mass. Above it, the green curve delimiting singularities with and without cosmological horizon cannot be drawn: either it is not defined, or it lies inside the inaccessible region (complex metric). \\

To confirm that \eqref{horizon greater than singularity alpha2=1/3} can be seen as a WGC bound, one can look at its behaviour when $H\to0$. To look at this limit, let us rewrite $r_h$, in the region where it is real, as 
\begin{equation}
    \label{rhorizon alpha 1/3}
    r_h=\frac{2}{\sqrt{3}H}\sin\left(\frac{\theta}{3}\right),
\end{equation}
where the angle $\theta$ is defined by $\sin(\theta)=({3\sqrt{3}}/{2})r_+H$ and $\cos(\theta)=\sqrt{1-({27}/{4})r_+^2H^2}$. In the limit $H\ll {1}/{r_+}$, one obtains 
\begin{equation}
    r_h=r_+ + H^2r_+^3+\mathcal{O}(r_+^3H^4).
\end{equation}
Looking at $r_h-r_-=0$, replacing $r_+$ and $r_-$ by their definition in function of $M$ and $Q$ \eqref{inverse relations constants physical parameters}, one can write the expansion of $Q$ in powers of $H$ as
\begin{equation}
\label{limit H to 0 for alpha2=1/3}
Q^2e^{2\alpha \phi_0}=\frac{4}{3}M^2+\frac{4^3}{3^4}M^4H^2+\mathcal{O}(M^6H^4).   
\end{equation}
In the limit $H\to0$, the bound given by \eqref{horizon greater than singularity alpha2=1/3} reduces to \eqref{dilatonicWGC}. \\

In conclusion, for $\alpha={1}/{\sqrt{3}}$, the study of horizons of these dilatonic black holes would rather suggest \eqref{horizon greater than singularity alpha2=1/3} as a WGC bound than \eqref{dilatonicWGC}.  

\subsubsection{$\alpha<\frac{1}{\sqrt 3}$}

As shown above, when $\alpha={1}/{\sqrt{3}}$, $r_+=r_-$ is no longer the black hole extremality condition, as it was for all cases with $\alpha>{1}/{\sqrt 3}$. In the following, we will see that this remains true for $\alpha<{1}/{\sqrt{3}}$.
\\

The first difference one can observe with respect to previous cases is the change in the behaviour of the derivative of $g_{00}(r)$ in a neighborhood of the singularity. Leading terms are given by
\begin{equation}
    \begin{cases}
  \partial_r (g_{00}) \underset{r\to r_-}{\sim}  -\frac{1-\alpha^2}{1+\alpha^2}\frac{r_-}{r^2}\left(1-\frac{r_+}{r}\right)\left(1-\frac{r_-}{r}\right)^{-2\frac{\alpha^2}{1+\alpha^2}}\qquad \mathrm{for}\; \alpha>\frac{1}{\sqrt{3}}\\
  \partial_r (g_{00}) \underset{r\to r_-}{\sim} 2\frac{\alpha^2}{1+\alpha^2}H^2r_-\left(1-\frac{r_-}{r}\right)^{-\frac{1-\alpha^2}{1+\alpha^2}} \qquad \mathrm{for}\; \alpha<\frac{1}{\sqrt{3}}.
    \end{cases}
\end{equation}
When $\alpha<{1}/{\sqrt{3}}$, the sign of the derivative at the vicinity of the singular surface $r=r_-$ is independent of $r_+$ and is always positive, with $\underset{r\to r_-^+}{\lim}g_{00}(r)=0^+$: $g_{00}$ always reaches $0$ from above. This, combined with the asymptotic value $g_{00}\to+\infty$ when $r\to \infty$, implies that the metric exhibits a horizon only if the parameters in $g_{00}$ are such that the function is decreasing in an interval to reach a negative minimum. In this situation, there are two horizons, coincident when the minimum of $g_{00}$ is $0$. This leads to a first conclusion:
\begin{itemize}
    \item In the parametric $(QH,MH)$ space,  a singularity surrounded only by a cosmological horizon can only appear on a curve, rather than in a portion of the plane as happened for all cases with $\alpha\ge\frac{1}{\sqrt 3}$.
\end{itemize}

Remember that, in contrast with the asymptotically flat case, here both $r_+>r_-$ and $r_+\leq r_-$ are now allowed. The method used above to investigate the limits between regions with different behaviours of the horizons was valid only for $r_+>r_-$ but can now be extended to all the situations. We proceed thus by using the function $F$ defined in \eqref{polynome for alpha2>1/3}, and its decomposition into $A$ and $B$. For $\alpha^2<{1}/{3}$, $B\to+\infty$ when $r\to r_-$. In a neighborhood of $r_-$ we always have $B(r)>A(r)$.

We notice that the condition $F(r_{0-})<0$ is not very illuminating in this case of $\alpha^2 < 1/3$. Indeed, for $r_+>r_-$, $F(r_{0-})$ is always negative while for $r_+\le r_-$, expanding \eqref{rzeroplus and rzerominus}, we have $r_{0-}\le r_-$, i.e. $r_{0-}$ lies inside the singular surface and $F(r_{0-})<0$ should not be considered.

One should recall here that, for any $\alpha\ne 0$, the solution is plagued by the appearance of the complex valued metric for $Q^2e^{2\alpha\phi_0}(1-\alpha^2)>M^2$. Sweeping the whole $r_+,r_-\in \mathbb R^+$ parametric space, we have no access to that region. All we can say here is that constraints from $F(r_{0-})$ do not appear in the region where the metric is defined real. For $\alpha=0$, the metric is real valued in the whole $(Q,M)$ plane: this allows us to verify that $F(r_{0-})=0$ matches the condition for the existence of naked singularities with a cosmological horizon in the RNdS metric found in \cite{Antoniadis:2020xso}.

Therefore:

\begin{itemize}
    \item The condition for the existence of black hole solutions is given by $F(r_{0+})>0$. When $F(r_{0+})=0$, the event and cosmological horizons coincide. $F(r_{0+})<0$ defines naked singularities with no cosmological horizon. 
\end{itemize}

There is a maximal mass, above which there is no black hole solution. This mass corresponds to the point where the curve defined by $F(r_{0+})=0$ crosses the line defined by $Q^2e^{2\alpha\phi_0}(1-\alpha^2)=M^2$. It is given by 
\begin{equation}
    M_{max}=\frac{1}{2\sqrt{2}H}\left(\frac{1-3\alpha^2}{2(1-\alpha^2)}\right)^{\frac{1-3\alpha^2}{2(1+\alpha^2)}}
\end{equation}

\begin{figure}[t]
    \centering
    \includegraphics[scale=0.7]{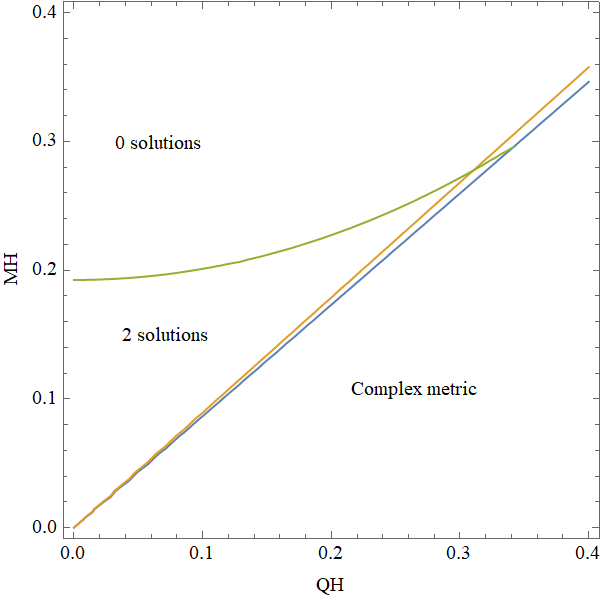}
    \caption{Number of horizons for $\alpha< {1}/{\sqrt{3}}$, here illustrated by the value $\alpha={1}/{2}$. The green curve represents $F(r_{0+})=0$ and gives an upper bound on the mass. The yellow line represents $Q^2=(1+\alpha^2)M^2$, that does not play anymore the same role for $\alpha< {1}/{\sqrt{3}}$. The blue one is $Q^2={M^2}/{(1-\alpha^2)}$. In the region between the green and the blue curves, cosmological and event horizons are present. Below the blue one, $r_+$, $r_-$ and the metric become complex valued. }
    \label{graphe alpha petit}
\end{figure}

The behaviour of the horizons for the asymptotically de Sitter metric for $\alpha < 1/ {\sqrt{3}}$ is described in figure \ref{graphe alpha petit} where we have taken, for an explicit illustrative example, $\alpha = 1/2$.

In figure \ref{graphe alpha petit}, the region of the $(QH,MH)$ plane with two horizons shows an upper bound represented by the green curve, $F(r_{0+})=0$. On the green curve, the event and cosmological horizons coincide. The lower bound is given by the blue curve, where $Q^2e^{2\alpha\phi_0}(1-\alpha^2)=M^2$ and the metric is on the verge of becoming complex. Plots of $g_{00}$ reveal that, on this line, the event horizon and the singularity are still far apart. Approaching this line from above (the black hole solution region), we see that the event horizon and the singularity get closer but never touch. Expected solutions with only the cosmological horizon seem to be hidden inside the inaccessible region. The maximal mass for the black hole, corresponding to the crossing point of the green and blue curves are given by $M_{max}H=2^{\frac{-8}{5}}3^{\frac{-1}{10}}\simeq0.3$.

We can compare with the RN-dS black hole solution studied in \cite{Romans:1991nq,Antoniadis:2020xso}, corresponding here to $\alpha=0$. In that case the $(QH,MH)$ plane shows a central region with three horizons surrounded by two regions with one horizon. One of them is attained, in parametric space, after the event horizon has reached the cosmological one, and is interpreted as a dS space-time causal patch eaten by the black hole. The other, related in \cite{Antoniadis:2020xso} to dS-WGC states, is beyond the locus of the coincidence of the inner and event horizons. For $0< \alpha < 1/ {\sqrt{3}}$, there are two horizons in a central region and zero horizons outside. Strictly speaking, $r=r_-$ is a zero of $g_{00}$ for all $\alpha<1$ but the inner horizon is traded for a singularity. The instability of Cauchy horizons may provide an additional motivation towards their identification. However, we have observed that for $\alpha \rightarrow 0$, while the two cosmological and event horizons tend to their corresponding surfaces in RN-dS, numerically the singularity $r_-$ seems not to coincide exactly with the inner horizon but lies slightly above: it is no more a singularity neither an horizon. In fact, in this limit, the expression of $r_-=r_-(Q,M)$, given in \eqref{inverse relations constants physical parameters}, takes the same form as the RN black hole inner horizon and becomes trivially this surface when $H\rightarrow 0$.

Finally, note that for $\alpha=0$ the parametric equations $F(r_{0 \pm})=0$ reproduce the relations separating the regions with different horizons in \cite{Antoniadis:2020xso}
\begin{equation}
\label{identification to the RN bounds}
    \begin{cases}
    F(r_{0-})=0 \underset{\alpha=0}{\Leftrightarrow}  M^2_-=\frac{1}{54l}[ l(l^2+36Q^2)-(l^2-12Q^2)^{\frac{3}{2}}]\\
    F(r_{0+})=0 \underset{\alpha=0}{\Leftrightarrow}  M^2_{+}=\frac{1}{54l}[ l(l^2+36Q^2)+(l^2-12Q^2)^{\frac{3}{2}}] 
    \end{cases}
\end{equation}
with $l=\frac{1}{H}$. It is $M^2_-$, and thus $F(r_{0-})$, that marks the transition between black holes and naked singularities with cosmological horizon. However, the solution of $F(r_{0-})=0$ can not be used for $0< \alpha < 1/ {\sqrt{3}}$, as the metric is complex in that region. Note that in all previous literature, because the asymptotically flat metric always shows a naked singularity before turning complex, this region was simply ignored.







\section{Dilatonic Anti-de Sitter Black Holes}

Changing the sign of the $H^2$ terms in the metric \eqref{metric alpha generic dS} allows to obtain a particular class of dilatonic asymptotically AdS black hole solutions. For completeness, we will investigate the phase space exhibiting the behaviour of the horizons as one varies $\alpha$, $M$ and $Q$  using the same method as for the de Sitter case.

We start by briefly recalling the Reissner Nordstr\"om AdS case as it will correspond to the $\alpha\to 0$ limit. The time component of the metric is $g_{00}(r)=-\left(1-\frac{2M}{r}+\frac{Q^2}{r^2}+H^2r^2\right)$. Its roots are given by those of the polynomial $G(r)\equiv H^2r^4+r^2-2Mr+Q^2$. They are two (degenerate in the extremal case) real positive roots as long as 

\begin{equation}
    M^2\ge \frac{1}{54}\left(36Q^2-\frac{1}{H^2}+\frac{\left(1+12 H^2 Q^2\right)^\frac{3}{2}}{H^2}\right).
    \label{bound RNAdS}
\end{equation}
The presence of horizons can be inspected through the study of the zeros of the function: 

\begin{equation}
F_{AdS}(r)\equiv r-r_+ +H^2r^3\left(1-\frac{r_-}{r}\right)^{\frac{3\alpha^2-1}{1+\alpha^2}}=-r\left(1-\frac{r_-}{r}\right)^{\frac{1-\alpha^2}{1+\alpha^2}}g_{00}(r).
\label{FAdS}
\end{equation}
This turns out to be much simpler to study than the corresponding dS function $F$. It is indeed straightforward to see that whenever $r_+<r_-$, $F_{AdS}>0$ for all $r\in\left[r_-,\infty\right[$ and so the function cannot have zeros. It will prove useful to split $F_{AdS}$ into the sum of the two contributions $A_{AdS}(r)\equiv r-r_+$, a straight line, and $B_{AdS}(r)\equiv -H^2r^3\left(1-\frac{r_-}{r}\right)^{\frac{3\alpha^2-1}{1+\alpha^2}}$ which is always negative. In this way, the problem is again recast in terms of the intersection points of $A_{AdS}$ and $B_{AdS}$. We split the discussion into three parts depending on the value of $\alpha$.
\\

\paragraph{\underline{
$\boldsymbol{\alpha^2>\frac{1}{3}}$}:} When approaching the singularity, $B_{AdS}$ goes to $0^-$. As a consequence, the curves defined by $A_{AdS}$ and $B_{AdS}$  have either one intersection point when $r_+ \geq r_- \,\left(A_{AdS}(r_-) \leq 0\right)$, or no intersections at all when $r_+<r_-\,\left(A_{AdS}(r_-)>0\right)$. We conclude that the discriminant between the black hole regime and the naked singularity is given by $r_+=r_-$ i.e. $Q^2e^{2\alpha\phi_0}=(1+\alpha^2)M^2$. \\
    
\begin{figure}[t]
    \centering
    \includegraphics[scale=0.5]{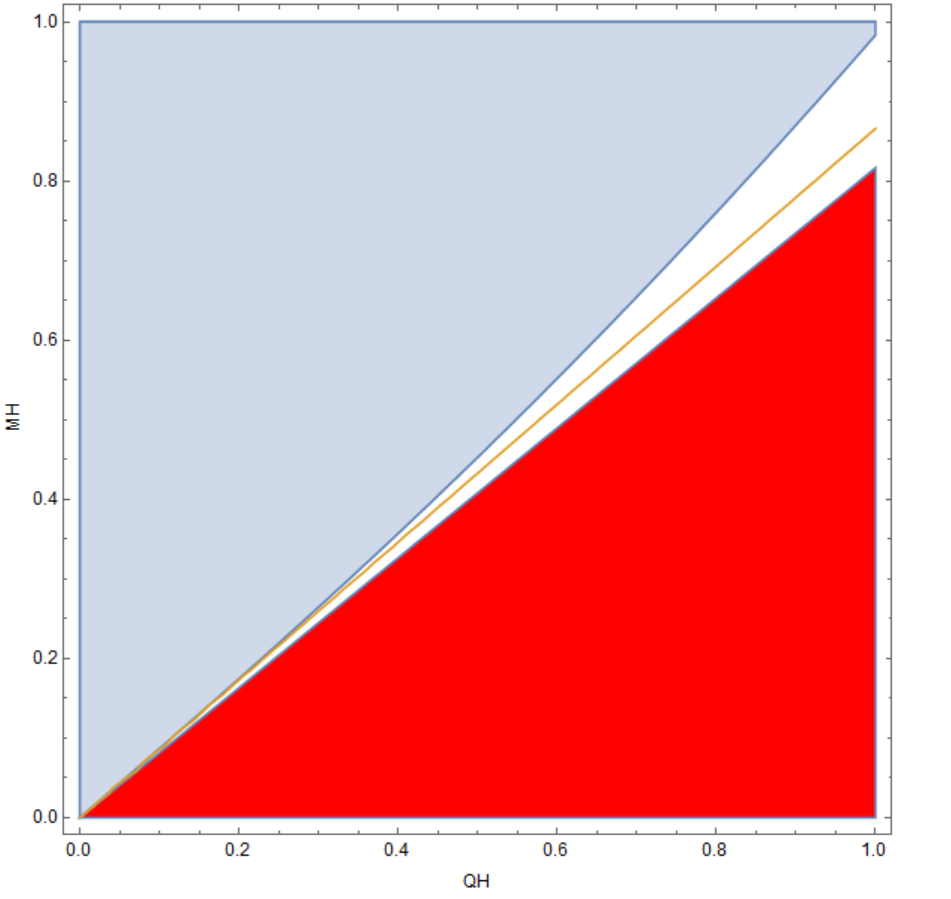}
    \includegraphics[scale=0.5]{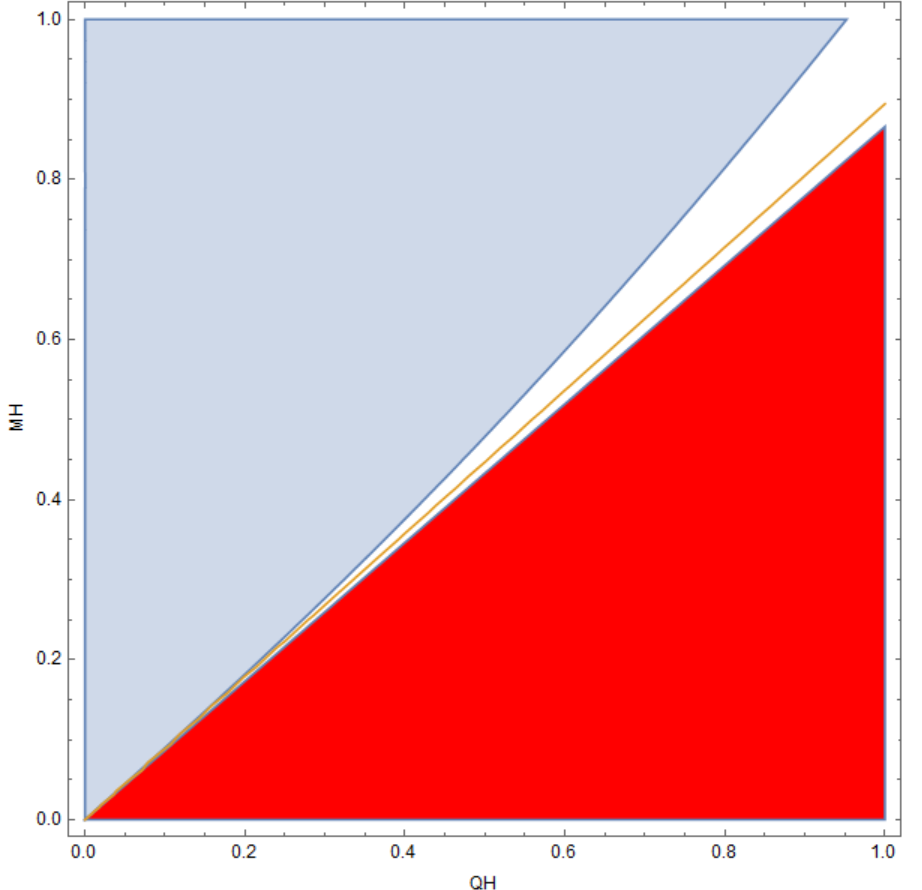}
    \caption{The dilatonic AdS black hole case. The {\it left panel} describes the phase diagram for the $\alpha^2={1}/{3}$, the  {\it right panel} shows the explicit example of $\alpha^2={1}/{4}$ to illustrate the situation for $\alpha^2<{1}/{3}$. In both cases, the blue region corresponds to black hole solutions, the yellow line is the flat-space discriminant between shielded and naked singularities (playing no role here but shown for comparison) and the red region is the inaccessible region where the metric becomes complex.}
    \label{AdS alpha^2=1/3 and less}
\end{figure}

\paragraph{\underline{$\boldsymbol{\alpha^2=\frac{1}{3}}$}:} $B_{AdS}$ does not depend on $r_-$: this is again related to a factorization in the metric as we have seen in the dS case. As such, $F_{AdS}$ now corresponds to the Schwarzschild-AdS polynomial $F_{AdS}(r)=r-r_+ +H^2r^3$, whose roots are given by

\begin{equation}
\begin{cases}
r_h=\frac{1}{H}\left(-\frac{\left(\frac{2}{3}\right)^{1/3}}{\left(9r_+H+\sqrt 3\sqrt{4+27r_+^2H^2}\right)^{1/3}}+\frac{\left(9r_+H+\sqrt 3\sqrt{4+27r_+^2H^2}\right)^{1/3}}{2^{1/3}3^{2/3}}\right) \\
r_{--}=\frac{1}{H}\left(\frac{\left(\frac{2}{3}\right)^{1/3}e^{-i\pi/3}}{\left(9r_+H+\sqrt 3\sqrt{4+27r_+^2H^2}\right)^{1/3}}-\frac{\left(9r_+H+\sqrt 3\sqrt{4+27r_+^2H^2}\right)^{1/3}e^{i\pi/3}}{2^{1/3}3^{2/3}}\right) \\
r_{--}^*=\frac{1}{H}\left(-\frac{\left(\frac{2}{3}\right)^{1/3}e^{i\pi/3}}{\left(9r_+H+\sqrt 3\sqrt{4+27r_+^2H^2}\right)^{1/3}}+\frac{\left(9r_+H+\sqrt 3\sqrt{4+27r_+^2H^2}\right)^{1/3}e^{-i\pi/3}}{2^{1/3}3^{2/3}}\right),
\end{cases}
\end{equation}
where $r_h$ is the horizon and $r_{--}$ and $r_{--}^*$ are two complex conjugate (non-physical) roots. Accordingly, the condition for the singularity to be shielded by the horizon is just $r_h > r_-$, which reads: 

\begin{align}
\label{bound AdS pour alpha2=1/3}
&\frac{\left(9\left(\hat M+\sqrt{\hat M^2-\frac{2}{3}\hat Q^2}\right)+\sqrt 3\sqrt{4+27\left(\hat M+\sqrt{\hat M^2-\frac{2}{3}\hat Q^2}\right)}\right)^{1/3}}{2^{1/3}\,3^{2/3}} \nonumber \\
&-\frac{\left(\frac{2}{3}\right)^{1/3}}{\left(9\left(\hat M+\sqrt{\hat M^2-\frac{2}{3}\hat Q^2}\right)+\sqrt 3\sqrt{4+27\left(\hat M+\sqrt{\hat M^2-\frac{2}{3}\hat Q^2}\right)}\right)^{1/3}} > \frac{4\hat Q^2}{3\left(M+\sqrt{M^2-\frac{2}{3}\hat Q^2}\right)}.
\end{align}
In the $(\hat Q,\hat M)$ space this gives a lower bound on the mass which is a little higher than the asymptotically flat case: $\hat Q^2>({4}/{3})\hat M^2$, as shown in figure \ref{AdS alpha^2=1/3 and less} (\textit{left} panel).  The $H\to 0$ limit of \eqref{bound AdS pour alpha2=1/3} is again given by \eqref{limit H to 0 for alpha2=1/3} with the change of sign in front of the $H^2$ term.\\

\paragraph{\underline{$\boldsymbol{\alpha^2<\frac{1}{3}}$}:} This is again the most intricate parametric region. Here, $B_{AdS}$ diverges to $-\infty$ when $r\to r_-$. For $r_+ \leq r_-$, since $A_{AdS}$ is positive for all $r\geq r_-$, and so is the difference $A_{AdS}-B_{AdS}$, no horizon can ever be present.

 On the other hand, when $r_+>r_-$ we have $A_{AdS}(r)<0$  for all $r\in[r_-,r_+[$, so the combination $A_{AdS}-B_{AdS}$ could result to be negative there. As $B_{AdS}$ is a concave function of $r$, two roots will be present when $A_{AdS}$ and $B_{AdS}$ intersect, collapsing to one when they are tangent, and zero otherwise. The region of parameters allowing the presence of two horizons, the black hole solution region, is obtained as in the dS case by solving the combined equations $F_{AdS}=0$ and $F_{AdS}'=0$. The solutions to this system are the same $r_{0\pm}$ found in \eqref{rzeroplus and rzerominus}\footnote{Writing $F_{AdS}'(r)$ with the prior $F_{AdS}(r)=0$, one sees that it is independent of the sign in front of the $H^2r^3$ term and gives the same equations as in the dS case, $F'(r)=0$ with the prior $F(r)=0$.}. In the $(r_+H,r_-H)$ plane, $F_{AdS}(r_{0+})$ is always null or positive, leading to no constraint in practice. As a consequence:

\begin{itemize}
\item The condition for the singularity to be shielded can be simply expressed as $F_{AdS}(r_{0-})\le 0$, with the equality being satisfied by extremal solutions with coincident horizons. 
\end{itemize}
This leads to a lower bound on the mass, that lies above the flat-space one ($Q^2e^{2\alpha\phi_0}=(1+\alpha^2)M^2$), as illustrated in the example of figure \ref{AdS alpha^2=1/3 and less} (\textit{right} panel). 

For $\alpha^2={1}/{3}$, the lower bound on the mass coincides with the one obtained above by simply requiring $r_h-r_-\ge 0$,  the set of curves are continuously connected. The presence of black holes with two horizons is a new characteristic that was not present for $\alpha\ge{1}/{\sqrt{3}}$. 

As in the dS case, we verify again the equivalence between the limit $\underset{\alpha\to 0}{\lim}F_{AdS}(r_{0-})\le 0$ and \eqref{bound RNAdS}.

Note that the singularity at $r=r_-$ changes its nature: from a space-like one, as it happens when it is behind the single $\alpha^2>{1}/{3}$ horizon, to a time-like one. This is dictated by the derivative of $g_{00}$ that diverges now to $-\infty$ for $r\to r_-^+$. The $\alpha^2<{1}/{3}$ AdS black holes are the only ones where the singularity can be avoided: $r=r_-$ is not in the future light-cone of all the observers that crosses the horizons, $g_{00}$ becomes time-like again before reaching it, as already observed by \cite{Goto:2018iay}. Of all the setups we studied, this is the only case where $\underset{r\to r_-^+}{\lim}g_{00}<0$ in a black hole parametric region. \\

This is similar to what happens in the Reissner Nordstrom AdS metric for the $r=0$ singularity. Varying $\alpha$, starting with $\alpha^2>{1}/{3}$, we encounter at $\alpha^2={1}/{3}$ a transition from Schwarzschild-AdS like black holes, with only one horizon and a space-like singularity, to Reissner-Nordstrom AdS like ones, with two horizons and a time-like singularity. For all $\alpha \neq 0$, the singularity at $r=r_-$ resembles here  the singularity at $r=0$ of $\alpha = 0$. Note that we do not encounter the issue of a complex metric, in contrast with the dS case, as the naked singularity bound is reached for values $M^2>(1-\alpha^2)Q^2$.\\

 It can be interesting to look at the first correction to the flat space-time condition $r_-=r_+$ for small $H$. For $H=0$, $F(r_{0-})=0$ reduces to $r_{0-}=r_+$ which is equivalent to $r_-=r_+$.\\
In order to find the first term in the expansion in $H$, we set 
\begin{equation}
\label{expansion r- ads}
r_-=r_+ +cr_+^{\gamma+1}H^{\gamma}+o(r_+^{\gamma+1}H^{\gamma}),    
\end{equation} where the constants $c$ and $\gamma$ have to be fixed. 

From \eqref{expansion r- ads}, it is possible to express $r_{0-}$ as $r_{0-}=r_+ + \frac{(1+\alpha^2)}{2(1-\alpha^2)}cr_+^{\gamma+1}H^{\gamma}+o(r_+^{\gamma+1}H^{\gamma})$. Requiring $F(r_{0-})=0$ at first order gives:
\begin{equation}
\gamma=\frac{1+\alpha^2}{1-\alpha^2} \qquad {\rm and} \qquad \frac{1+\alpha^2}{2(1-\alpha^2)}c+\left[\frac{3\alpha^2-1}{2(1-\alpha^2)} c\right]^{\frac{3\alpha^2-1}{1+\alpha^2}}=0.   
\label{c-def}
\end{equation}
For $\alpha<{1}/{\sqrt{3}}$, $c$ is single valued, negative, with the limits $c\to-2$ when $\alpha\to 0$ and $c\to-1$ for $\alpha\to{1}/{\sqrt{3}}$.
Plugging the relation between $(r_+,r_-)$ and $(Q,M)$ given by \eqref{inverse relations constants physical parameters} in \eqref{expansion r- ads}, we can look for the corresponding relation $Q^2e^{2\alpha\phi_0}=(1+\alpha^2)M^2+b M^{2+\delta}H^{\delta}$, that defines the boundary of the black hole region. It is possible to determine the constants  $\delta$ and $b$:  $\delta=\gamma$ and $b=\alpha^2(1+\alpha^2)^{\frac{2}{1-\alpha^2}}c$. Thus, the constraint $F(r_{0-})=0$ can be expanded for $H\to0$ as 
\begin{equation}
    Q^2e^{2\alpha\phi_0}=(1+\alpha^2)M^2+\alpha^2(1+\alpha^2)^{\frac{2}{1-\alpha^2}}c\, M^{\frac{3-\alpha^2}{1-\alpha^2}}H^{\frac{1+\alpha^2}{1-\alpha^2}}+o(M^{\frac{3-\alpha^2}{1-\alpha^2}}H^{\frac{1+\alpha^2}{1-\alpha^2}}).
\end{equation}
We see that for $\alpha={1}/{\sqrt{3}}$, it reduces to 
\begin{equation}
    Q^2e^{2\alpha\phi_0}=\frac{4}{3}M^2-\frac{4^3}{3^4}M^4H^2+o(M^4H^2).
\end{equation}
It is the same equation as for the dS case, with a difference of sign.
For $\alpha\to0$, the power of H tends to $1$, but the coefficient in front vanishes. This is coherent with \cite{Romans:1991nq}, since there is no linear term in the expansion for small $H$.








\section{Thermodynamics}
\label{thermodynamics section}

The analysis of the existence of horizons shows that above and below the value $\alpha=1/\sqrt 3$ both the dS and AdS black holes have different properties. We have seen these differences in the behaviours of $g_{00}$ and its first derivative $\partial_r g_{00}$. In this section, we shall analyse them using thermodynamic quantities of the black holes \footnote{For the physics of thermodynamic quantities in de Sitter space-time, we refer the reader to \cite{Witten:2001kn}.}. 
\\

The Hawking temperature of the black holes is given by

\begin{align}
\label{temperature}
T=-\frac{\partial_rg_{00}}{4\pi}\Bigg|_{r=r_h}=\frac{1}{4\pi}& \Bigg[ \frac{r_+}{r_h}\left(1-\frac{r_-}{r_h^2}\right)^{\frac{1-\alpha^2}{1+\alpha^2}}+\frac{1-\alpha^2}{1+\alpha^2}\left(1-\frac{r_+}{r_h}\right)\left(1-\frac{r_-}{r_h}\right)^{-\frac{2\alpha^2}{1+\alpha^2}}\frac{r_-}{r_h^2} \nonumber \\
& \,\,\mp2H^2r_h\left(1-\frac{r_-}{r_h}\right)^{\frac{2\alpha^2}{1+\alpha^2}} \mp2\frac{\alpha^2}{1+\alpha^2}H^2r_-\left(1-\frac{r_-}{r_h}\right)^{-\frac{1-\alpha^2}{1+\alpha^2}} \Bigg],
\end{align}
where $r_h$ is the radial coordinate of the event horizon. The upper (lower) sign refers to the dS (AdS) solution. The above expression could be simplified using the relation between $r_+, r_-$ and $r_h$ given by $g_{00}(r_h)=0$, but it is more useful for our purposes to keep it in this form. Note that the second term in \eqref{temperature} does not vanish identically as now $r_h\ne r_+$ in general. We write the Hawking-Beckenstein black hole entropy proportional to the horizon area as

\begin{equation}
\label{entropy black holes}
    S=\pi r_h^2\left(1-\frac{r_-}{r_h}\right)^{\frac{2\alpha^2}{1+\alpha^2}},
\end{equation}
 Note that this is the same as in asymptotically flat space and vanishes when $r_h\to r_-$ for any $\alpha \ne 0$. \\

For $\alpha>1/\sqrt{3}$ extremal solutions are defined to have $r_h\to r_-=r_+$. The temperature diverges in the limit $r_h\to r_-$ when $\alpha\ne 1$ and we distinguish two cases.

\begin{itemize}
    \item For $\alpha>1$, the first and the second term of \eqref{temperature} lead the divergence. The temperature goes as
    
    \begin{equation}
    T\underset{r_h\to r_-}{\sim}\frac{1}{4\pi r_h} \left(1+\frac{1-\alpha^2}{1+\alpha^2}\right)\left(1-\frac{r_-}{r_h}\right)^{\frac{1-\alpha^2}{1+\alpha^2}}.
    \end{equation} 
    
    \item For $1/\sqrt{3}<\alpha<1$, the divergence is lead by the last term ($H\neq 0$)  
    \begin{equation}
    \label{temperature under 1}
    T\underset{r_h\to r_-}{\sim}\frac{1}{2\pi}\frac{\alpha^2}{1+\alpha^2}H^2r_-\left(1-\frac{r_-}{r_h}\right)^{-\frac{1-\alpha^2}{1+\alpha^2}}
    \end{equation} 
\end{itemize}
For $\alpha=1$, the temperature is finite and reads

\begin{equation}
\label{temperature alpha=1}
T=\frac{1}{4\pi}\left(\frac{1}{2M}\mp2MH^2\right)    
\end{equation}
In this case, with the extremality condition $D=M$ and the necessary requirement for the singularity to be smaller than the Hubble radius in the dS case, $4D^2H^2\le 1$, the expression \eqref{temperature alpha=1} is always positive (or null). Actually, this points out a rather interesting property: the extremal solution with the singularity of the same size as the Hubble horizon (denoted by the point where the green, yellow and red curve of figure \ref{fig:my_label} meet) has a null temperature. It is thus a trivial endpoint of Hawking evaporation. This is a not the case in the asymptotically flat metric where $T={1}/{8\pi M}$ and one questioned whether the extremal solutions are endpoints of Hawking evaporation or not \cite{Holzhey:1991bx}. Note that the finiteness of $T$ comes from the vanishing of the exponent $\left|{(1-\alpha^2)}/{(1+\alpha^2)}\right|$ in the continuous limit $\alpha \to 1$.

The divergence of the temperature was discussed in \cite{Holzhey:1991bx} for the asymptotically flat $\alpha>1$ case. The divergence for $1/\sqrt{3}<\alpha<1$ is new and entirely due to the presence of a non-vanishing cosmological constant. For the presence of horizons, the $1/\sqrt{3}<\alpha<1$ and $\alpha>1$ black holes share the same properties and are not much sensitive to the value of $H$ when they approach extremal solutions. The temperature, however, has a different form in the two cases and shows a dependence on $H$.\\

When $0<\alpha\le1/\sqrt 3$, extremal solutions no longer have $r_+=r_-$. We distinguish two cases.

\begin{itemize}
    \item For $\alpha=1/\sqrt{3}$, the extremality condition $r_h=r_-$ is reached when $F(r_{0-})=0$ (defined in \eqref{polynome for alpha2>1/3} and \eqref{rzeroplus and rzerominus})  or $F_{AdS}(r_{0-})=0$ (defined in \eqref{FAdS}). The temperature of such black holes is 
    
    \begin{equation}
    T\underset{r_h\to r_-}{\sim}\frac{1}{8\pi }\frac{r_-}{r_h^2}\left(1-\frac{r_+}{r_h}\mp H^2r_h^2 \right)\left(1-\frac{r_-}{r_h}\right)^{-\frac{1}{2}}=0, 
    \end{equation}
    thus vanishes for the extremal solution as the factor inside the first parenthesis corresponds to $g_{00}(r_h)$ and is identically null. If one were to blindly take the limit $\alpha\to 1/\sqrt{3}$ of \eqref{temperature under 1}, the temperature of extremal solutions would seem to diverge with an exponent $1/2$.  This shows a discontinuity in the $\alpha$-dependence of such exponent that can be traced back to the factorization in the metric and the loss of the extremality condition $r_+=r_-$. 
    
    \item When $0<\alpha<1/\sqrt 3$, we have seen in both the dS and AdS cases that  $r_h=r_-$ is never reached for different reasons.\\ 
    
    In the dS case, extremal solutions could not be defined within the domain of a real valued metric.  \\

    In the AdS case, the black holes on the verge of exposing a naked singularity have the event horizon coincident with their inner horizon. As these do not coincide with the singularity, the temperature does not diverge anymore but vanishes as $\partial_r g_{00}(r_h)=0$.
    
\end{itemize}

A simple interpretation of the behaviour of the temperature is as follows. The $\alpha\to \infty$ and $\alpha\to 0$ limits of such black holes are given by Schwarzschild and Reissner-Nordstr\"om black holes, respectively, with temperatures given by 
\begin{equation}
    T_{Sc}=\frac{1}{8\pi M}, \qquad  T_{RN}=\frac{1}{2\pi}\frac{\sqrt{M^2-Q^2}}{\left(M+\sqrt{M^2-Q^2}\right)^2}
\end{equation}
in asymptotically flat space. In that case, the dilatonic black holes have a temperature

\begin{equation}
    T=\frac{1}{4\pi r_+}\left(1-\frac{r_-}{r_+}\right)^{\frac{1-\alpha^2}{1+\alpha^2}},
\end{equation}
diverging for $\alpha>1$, finite for $\alpha=1$, and vanishing for $0<\alpha<1$ extremal solutions. We retrieve a vanishing temperature for small $\alpha$, here $\alpha=0$, as the extremality condition reads now $M=Q$. For large values of $\alpha$,  the extremality condition obtained by identification of the horizon $r_h(=r_+)$ with  the  singularity $r_-$, would formally correspond in  the Schwarzschild case to put the horizon at the origin i.e. formally take the limit $M$ tends to $0$ in the black hole solution which in turn leads to a  divergent temperature. 

This remains true also in asymptotically (A)dS space-time. For large $\alpha$, we consider  the temperature of the Schwarzschild (A)dS black hole

\begin{equation}
\label{temperature S(A)dS}
    T_{Sc}=\frac{1}{4\pi}\frac{1\mp3H^2r_h^2}{r_h}, 
\end{equation}
where $r_h$ is the radius of the event horizon \footnote{Note, that in the dS case, the singularity is shielded as long as $M^2H^2<1/27$, while in the AdS case it is for any mass $M>0$. We do not discuss issues related to thermal equilibrium of observers in the region between the event and cosmological horizons in asymptotically dS space \cite{Gibbons:1977mu}. We restrict to the small $H$ limit where the two are far away.}. The extremality corresponds again to a formal $r_h \rightarrow 0$ limit, obtained when $M\to0$ in \eqref{temperature S(A)dS}. This leads to a divergent temperature.
For small $\alpha$, we consider instead the Reissner-Nordstr\"om (A)dS black holes, and $T$ has a similar form \cite{Romans:1991nq}

\begin{equation}
    T_{RN}=\frac{1}{4\pi}\frac{1-\frac{Q^2}{r_h^2}\mp3H^2r_h^2}{r_h}, 
\end{equation}
where $r_h$ indicates again the radius of the event horizon but defined by a different metric. The region of validity for the dS and AdS solutions are given in \eqref{identification to the RN bounds} and \eqref{bound RNAdS}, respectively. Extremality is obtained then by taking the lower mass bound for which the temperature vanishes.  \\

In this picture, the transition between diverging and vanishing temperatures of extremal dilatonic black holes could be then seen as the thermal footprint of a transition from a Schwarzschild-like behaviour to a Reissner-Nordstr\"om like one. Such transition happens for $\alpha=1$ in asymptotically flat space, where the temperature is equal to $T=\frac{1}{8\pi M}$, and for $\alpha=1/\sqrt{3}$ in asymptotically (A)dS space, where $T=0$.  \\

For the AdS black holes, one can observe a peculiar behaviour of the entropy formula applied to the extremal solution. It trivially vanishes above $\alpha=1/\sqrt 3$ as one has a naked singularity. It is finite for $\alpha<1/\sqrt 3$, increasing as $\alpha \to 0$: there, the extremal condition corresponds to the coincidence of two (non-singular) horizons. \\






\section{Test particles in charged dilatonic black hole metric}

The weak gravity conjectures, for abelian gauge symmetries, dilatonic or scalar interactions, have been formulated as constraints on the non-relativistic and, often but not always, large distance interactions between charged states. These states can be elementary in the theory, but also solitonic as D-branes. Accordingly, we wish to study non-relativistic, large distance, leading interactions between the charged black holes. The latter, separated by very large distances and interacting through gravitons, gauge bosons and scalar fields with large wavelengths compared to their typical size, i.e. their horizon radius, look like point particles. One challenge for the point-like particle description of the interactions is to identify here the expression of the scalar coupling and associated scalar charge of these states.

It appears instructive to first consider the simpler case of a test particle submitted to the forces generated by a black hole. Also, taking in our computations the limit $H=0$, allows to compare with the available results of explicit amplitude computation.

\subsection{Large distance action of the dilatonic black holes on a test particle}

In our effective theory description, the scalar charge of a point-like particle  with respect to the dilaton $\phi$ appears encoded in the field-dependent mass $m(\phi)$. This in turn will be translated into a three-point coupling in a field theory context, as we shall discuss later. The action for the motion of a test particle of mass $m(\phi)$ and charge $q$ in the black hole geometry is given by
\begin{equation}
\label{point particle action}
    S_m=\int \mathrm{d}\tau \left(- m(\phi)\sqrt{-g_{\mu \nu}\Dot{x}^{\mu}\Dot{x}^{\nu}}+\sqrt{4\pi G}gqA_{\mu}\Dot{x}^{\mu}\right),
\end{equation}
where $x^{\mu}$ represent the particle's coordinates and the dot indicates a derivative with respect to the proper time $\tau$. The last term is the coupling to the abelian gauge field $A_\mu$, with a gauge coupling constant $g$. The mass $m$ and the charge $q$ are in geometrized units for consistency. 
The geodesic equations are

\begin{equation}
-m(\phi)\left(\ddot x^\mu+\Gamma^\mu_{\nu\rho}\dot x^\nu\dot x^\rho\right)+\sqrt{4\pi G}gqF^\mu_{\,\,\rho}\dot x^\rho-\frac{d m(\phi)}{d\phi}\left(\dot x^\mu \dot x^\rho\partial_\rho \phi-\dot x^\rho \dot x_\rho\partial^\mu\phi\right)=0,
\end{equation}
where the $\Gamma$s denote the Christoffel symbols and $F_{\mu \rho}$ is the gauge field strength. Here, we are mainly interested in the last term, interpreted as a scalar force. 

To study the motion of the test particle in the space-time defined by the metric \eqref{metric alpha generic dS}, we first rewrite the Lagrangian as

\begin{equation}
\mathcal L=-m(\phi)\sqrt{f(r)\dot t^2-\frac{\dot r^2}{f(r)}-r^2g(r)\dot\theta^2-r^2g(r)\sin^2{\theta}\,\dot\varphi^2}-\frac{e^{2\alpha\phi_0}q{Q}}{r}\dot t, 
\label{Lagrangian rewritten}
\end{equation}
where the gauge field was chosen as $A=\left(-\frac{gQ}{\sqrt{4\pi G }r},0,0,0\right)$ and the gauge coupling is now $g=e^{\alpha\phi_0}$. In \eqref{Lagrangian rewritten}, we have introduced two functions $f$ and $g$:

\begin{equation}
    \begin{cases}
    f(r)\equiv  \left(1-\frac{r_+}{r}\right)\left(1-\frac{r_-}{r}\right)^{\frac{1-\alpha^2}{1+\alpha^2}}\mp H^2r^2\left(1-\frac{r_-}{r}\right)^{\frac{2\alpha^2}{1+\alpha^2}} \\
    g(r)\equiv \left(1-\frac{r_-}{r}\right)^{\frac{2\alpha^2}{1+\alpha^2}},
    \end{cases}
\end{equation}
where the $\mp$ signs depend on whether we consider an asymptotically dS or AdS space-time, respectively.  

We shall start by taking $H=0$, corresponding to the asymptotically flat space-time solution, and discuss the $H\ne 0$ case in a second moment.

The spherical symmetry allows us to restrict the analysis to the equatorial plane $\theta=\frac{\pi}{2}$. The two Killing vectors $\partial_t$ and $\partial_\varphi$ correspond to two constant conserved quantities $E$ and $L$, proportional to the energy and angular momentum as measured at infinity, respectively. They are given by:
\begin{equation}
\begin{cases}
\label{energy and angular mom}
E=-\frac{\partial \mathcal L}{\partial \dot t}=m(\phi)f(r)\dot t+\frac{
e^{2\alpha\phi_0}q Q }{r} \\
L=\frac{\partial \mathcal L}{\partial \dot\varphi}=m(\phi)r^2g(r)\dot\varphi ,
\end{cases}
\end{equation}
In \eqref{energy and angular mom} we have used the normalization $g_{\mu \nu}\Dot{x}^{\mu}\Dot{x}^{\nu}=-1$ whose explicit form reads 

\begin{equation}
\label{Geodesic in GHS with angular momentum}
-f(r)\dot t^2+\frac{\dot r^2}{f(r)}+r^2g(r)\dot \varphi^2=-1,
\end{equation}
or, using \eqref{energy and angular mom},

\begin{equation}
\label{Geodesic in GHS without angular momentum}
-\frac{1}{m^2(\phi)f(r)}\left(E-\frac{e^{2\alpha\phi_0}q Q}{r}\right)^2+\frac{\dot r^2}{f(r)}+\frac{L^2}{m^2(\phi)r^2g(r)}=-1.
\end{equation}
Restricting to radial paths and null angular momentum $L$, this gives

\begin{equation}
 \left(\frac{\mathrm{d}r}{\mathrm{d}\tau}\right)^2=-f(r)+\frac{1}{m^2(\phi)}\left(E-\frac{e^{2\alpha\phi_0}q Q}{r}\right)^2 .  
\end{equation}
After putting the equation in the form $\frac{1}{2}\left(\frac{\mathrm{d}r}{\mathrm{d}\tau}\right)^2+V_{\mathrm{eff}}(r)=0$, one can read the forces from the ${1}/{r}$ coefficient in $V_{\mathrm{eff}}(r)$, the Newtonian approximation being recovered at large distances ($r\gg r_-$). In this limit, using \eqref{metric alpha generic dS}, the leading order of $f$ and $m$ are
\begin{align}
    &f(r)=1-\frac{1}{r}\left(r_{+}+\frac{1-\alpha^2}{1+\alpha^2}r_{-}\right)+\mathcal{O}\left(\frac{1}{r^2}\right) \\
    &m^2(\phi)=m^2\left(\phi_0-\frac{\alpha}{1+\alpha^2}\frac{r_-}{r}+\mathcal{O}\left(\frac{1}{r^2}\right)\right)=m^2(\phi_0)-\frac{\mathrm{d}m^2}{\mathrm{d}\phi}\Bigg|_{\phi_0}\frac{\alpha}{1+\alpha^2}\frac{r_-}{r}+\mathcal{O}\left(\frac{1}{r^2}\right).
\end{align}
Together with the identification in \eqref{relations constants physical parameters}, this gives  
\begin{equation}
\frac{1}{2}\left(\frac{\mathrm{d}r}{\mathrm{d}\tau}\right)^2=\frac{\frac{E^2}{m_0^2}-1}{2}+\frac{M}{r}+\frac{E^2}{m_0^2}\frac{\frac{m'_0}{m_0}D}{r}-\frac{E}{m_0}\frac{e^{2\alpha\phi_0}\frac{q}{m_0} Q}{r}+\mathcal{O}\left(\frac{1}{r^2}\right) \label{acceleration MD}   
\end{equation}
where the $'$ stands for the derivative with respects to $\phi$ and the subscripts $0$ denote quantities evaluated at $\phi=\phi_0$. In \eqref{acceleration MD}, $M$, $Q$ and $D$ are the mass, the charge and scalar charge of the black hole expressed in geometrized units. Note that $D$ is a secondary charge related to $M$ and $Q$ in \eqref{definition scalar charge}. The ${E}/{m_0}$ factors should be seen as relativistic corrections, intrinsically present in the GR framework, and important for ${v}/{c} \sim 1$. At leading order, the non-relativistic potential $V_{pp}$ felt by a point-particle takes then the form 

\begin{equation}
\label{effective potential motion in GHS metric}
V_{pp}(r)\equiv m_0V_{\mathrm{eff}}(r)=-\frac{m_0M +m'_0D-e^{2\alpha\phi_0}q Q}{r}+\mathcal{O}\left(\frac{1}{r^2}\right).    
\end{equation}
We have therefore shown that the forces felt by a test particle match the result for the non-relativistic limit of the $2\to 2$ scattering amplitude between this test particle and a state with the gauge charge, scalar charge and mass of the black hole. \\

Before turning to the question of how to extend this picture to the case of scattering of two point-like dilatonic black holes, we give here the final expression for the second order term in the expansion of $V_{\mathrm{eff}}(r)$ 

\begin{equation}
\begin{split}
 V^{(2)}_{\mathrm{eff}}(r)= -\frac{1}{2r^2}\Bigg[ &\frac{e^{4\alpha\phi_0}q^2}{m_0^2} Q^2 -(1-\alpha^2)\left(e^{2\alpha\phi_0} Q^2-D^2\right) -\frac{1}{2}\frac{E^2}{m^2_0}D^2\frac{{{m''}_0}^2}{m^2_0}  \big. \\ 
 &\left. +\frac{E^2}{m_0^2}\frac{1+\alpha^2}{\alpha}D^2\frac{m'_0}{m} -4\frac{E}{m_0}e^{2\alpha\phi_0}\frac{q}{m_0} Q D \frac{m'_0}{m_0}
 +4\frac{E^2}{m_0^2}D^2\left(\frac{m'_0}{m_0}\right)^2 \right],   
\end{split}   
\label{the r2 expansion}
\end{equation}
This generalizes the  $1/r^2$ term one finds in the Reissner-Nordstr\"om case:

\begin{equation}
V^{(2)}_{\mathrm{eff}}(r)=-\frac{1}{2r^2}\left(\frac{q^2}{m^2}Q^2- Q^2\right)    
\end{equation}
that is recovered in the limit $\alpha\to 0$ (thus $D\to 0$), giving a non vanishing contribution even for purely radial motion. Actually, in the dilatonic case, the expansion of the various terms formally gives contributions to the different orders in $1/r$.

For generic paths with $L\ne 0$, the $1/r^2$ term gets contributions both from \eqref{the r2 expansion} and from the angular momentum term 
\begin{equation}
 \frac{1}{m^2(\phi)}\frac{L^2}{r^2}\frac{f(r)}{g(r)} \qquad {\rm with} \qquad \frac{f(r)}{g(r)}=\left(1-\frac{r_+}{r}\right)\left(1-\frac{r_-}{r}\right)^{\frac{1-3\alpha^2}{1+\alpha^2}}   
\end{equation}
The explicit expression of the angular momentum-related potential term is 
\begin{align}
\frac{1}{m^2}\frac{L^2}{r^2}\frac{f}{g}=\frac{L^2}{m_0^2r^2}&\Bigg[1+\frac{2D\frac{m'_0}{m_0}-2M+2\alpha D}{r}\nonumber \\
&+\frac{1}{r^2}\bigg[\frac{1+\alpha^2}{\alpha}\frac{m'_0}{m_0}D-\frac{{m^2}''_0}{m^2_0}\frac{D}{2}+2\left(\frac{m'_0}{m_0}\right)^2+(1-3\alpha^2)e^{2\alpha\phi_0} Q^2 +4(1+\alpha^2)\frac{m'_0}{m}D^2 \nonumber\\
&\qquad \,\,\,-2(1+\alpha^2)(1-3\alpha^2)D^2
-2\frac{m'_0}{m_0}D M -2\frac{1-\alpha^2}{\alpha}\frac{m'_0}{m_0}D\bigg]+\mathcal O\left(\frac{1}{r^3}\right)\Bigg]. 
\end{align}
In the limit $\alpha\to 0$ this reduces to 
\begin{equation}
 \frac{L^2}{m^2r^2}\left(1-\frac{2M}{r}+\frac{Q^2}{r^2}\right)   
\end{equation} 
An important difference for $\alpha \neq 0$, for the overall potential sourced by both $V_\mathrm{eff}$ and the angular momentum, is that the sub-leading contributions are parts of a formally infinite expansion in powers of ${r_-}/{r}$, where $r=r_-$ is the location of the singularity. As we approach $r_-$, higher orders become important and the whole expansion needs to be taken into account. Contrary to Schwarzschild or Reissner-Nordstr\"om solutions, there is no fixed-order dominant term whose sign determines whether the overall effective potential is attractive or repulsive around the singularity. \\

In the presence of a non-zero cosmological constant, the computation leading to \eqref{Geodesic in GHS with angular momentum} and \eqref{Geodesic in GHS without angular momentum} is unchanged but with the additional  $\mp H^2r^2\left(1-\frac{r_-}{r}\right)^{\frac{2\alpha^2}{1+\alpha^2}}$ in $f(r)$. At large $r\gg r_-$, this can be expanded to give 

\begin{align}
\mp H^2r^2\left(1-\frac{r_-}{r}\right)^{\frac{2\alpha^2}{1+\alpha^2}}\underset{r\gg r_-}{\sim}  \mp &\left(H^2r^2-2\alpha D H^2 r+(\alpha^2-1)D^2H^2+\frac{2}{3}\frac{\alpha^2-1}{\alpha}\frac{D^3H^2}{r} \right. \nonumber \\   &\left. \,\,+\frac{1}{6}\frac{(\alpha^2-1)(\alpha^2+3)}{\alpha^2}\frac{D^4H^2}{r^2} +\mathcal O\left(\frac{r_-^5}{r^5}\right)\right).
\end{align}
 The result is an additional contribution to the ${1}/{r}$ potential, trivially vanishing for $H\to 0$, $\alpha\to 0$ and $\alpha=1$.  Finally, the additional contribution to the angular momentum terms, given by $({L^2}/{m^2(\phi)r^2})({f(r)}/{g(r)})$ leads to $\mp {L^2H^2}/{m^2_0}$, at leading order.

\subsection{Forces between two point-like states with the black holes charges }

We will investigate now the interaction between two point-like states both describing the black hole type solutions. These states will be characterized by their mass, their charge and their coupling to the dilaton $\phi$. We will pursue this description in region of parameters of the solution even beyond the extremal solution limit, therefore point-like states not corresponding to black holes anymore, in an attempt to get some indication of what happens where the metric becomes complex. \\

The question of how to associate the parameters of a dilatonic black hole to a particle state was addressed in \cite{Julie:2018lfp,Khalil:2018aaj} for the case of an asymptotically flat space-time solution. \\
The black hole parameters (say its ADM mass, gauge charge and scalar charge) are defined at infinity. As such, for a point particle to effectively describe this black hole, its charge $q$, mass $m(\phi)$ and first derivative $m'(\phi)$ observed at infinity must satisfy the conditions

\begin{equation}
\label{m,q,d relations julie}
\begin{cases}
m(\phi_0)=M=\frac{1}{2}\left(r_+ +\frac{1-\alpha^2}{1+\alpha^2}r_- \right) \\
q=Q=\sqrt{\frac{r_+r_-}{1+\alpha^2}}e^{-\alpha\phi_0} \\
m'(\phi)\big|_{\phi_0}=D=\frac{\alpha}{1+\alpha^2}r_-,
\end{cases}
\end{equation}
where $\phi_0$ is the asymptotic value of $\phi$ at infinity.

In order to obtain an explicit expression for the scalar charge/scalar coupling of the point-like black hole approximation, we express the relation \eqref{definition scalar charge} as:
\begin{equation}
D=\frac{\alpha}{1-\alpha^2}\left(M-\sqrt{M^2-(1-\alpha^2)Q^2e^{2\alpha\phi_0}}\right)    
\end{equation}
and consider that the point-like state lives in a region where $\phi_0\simeq \phi$, generated by the other (distant) black hole, and therefore has a coupling to the dilaton given by:

\begin{equation}
\label{differential equation mass Julie}
\frac{\mathrm{d}m}{\mathrm{d}\phi}=\frac{\alpha}{1-\alpha^2}\left(m(\phi)-\sqrt{m^2(\phi)-(1-\alpha^2)q^2e^{2\alpha\phi}}\right).    
\end{equation}
As was shown in \cite{Julie:2018lfp,Khalil:2018aaj}, the useful parameters to describe the scalar interactions are 

\begin{equation}
\gamma(\phi)\equiv\frac{\mathrm d}{\mathrm d\phi}\ln{m(\phi)} \quad\mathrm{and}\quad \beta(\phi)\equiv\frac{\mathrm d\gamma(\phi)}{\mathrm d \phi},  
\end{equation}
and the mass $m(\phi)$ can be expanded around a background value $\bar \phi$ as 

\begin{equation}
m(\phi)=m(\bar \phi)\left(1+\gamma(\bar \phi)(\phi-\bar \phi)+\frac{1}{2}\left(\gamma^2(\bar \phi)+\beta(\bar \phi)\right)(\phi-\bar \phi)^2+\mathcal O\left((\phi-\bar \phi)^3\right)\right).  
\label{mass expansion dilaton}
\end{equation}
Using \eqref{differential equation mass Julie} one obtains

\begin{equation}
\begin{cases}
\gamma(\phi)=\frac{\alpha}{1-\alpha^2}\left(1-\sqrt{1-(1-\alpha^2)\frac{q^2}{m^2(\phi)}e^{2\alpha\phi}}\right) \\
\beta(\phi)=\frac{\alpha^2}{1-\alpha^2}\frac{q^2e^{2\alpha\phi}}{m^2(\phi)}\left(1-\frac{\alpha^2}{\sqrt{1-(1-\alpha^2)\frac{q^2}{m^2(\phi)}e^{2\alpha\phi}}}\right).
\end{cases}    
\end{equation}
With these formulae at hand, we can now extend the analysis of one black hole and a test particle case to the present case with two black holes.\\

For $\alpha=1$, it is easy to see that an explicit solution is $m(\phi)=\sqrt{\mu^2+{q^2e^{2\phi}}/{2}}$, where $\mu$ is an integration constant. It is useful for the discussion below to recall that geometrized units have been used so far and that the $\phi$ field here is dimensionless (see \eqref{physical fields}). In terms of physical quantities, this translates into

\begin{equation}
\label{mass case alpha=1}
m(\phi)=\sqrt{\mu^2+M_P^2q^2e^{\sqrt{2}\frac{\phi}{M_P}}},
\end{equation}
where, although we have used the same notation for simplicity, the quantities should now be understood to be the physical ones. \\

The tree-level $t$-channel contribution to the $2\to2$ scattering amplitude of 2 such states with same charge $q$ and mass $m(\phi)$ reads\footnote{The two states being the same, they have the same asymptotic value of $\phi$ thus $\bar \phi = \phi_0$. We however keep the bar notation here.}:

\begin{equation}
    \mathcal A=\frac{4m^2}{t}\left(q^2e^{\sqrt 2\frac{\phi}{M_P}}-\left(\partial_\phi m\right)^2-\frac{1}{2}\frac{m^2}{M_P^2}\right)\Bigg|_{\bar \phi}=\frac{4m^2}{t}\left(q^2e^{\sqrt 2\frac{\phi}{M_P}}-\frac{1}{2}\frac{M_P^2}{m^2}q^4e^{2\sqrt 2\frac{\phi}{M_P}}-\frac{1}{2}\frac{m^2}{M_P^2}\right)\Bigg|_{\bar \phi},
\end{equation}
where the bar from now on indicates quantities evaluated at the background value $\bar \phi$.
The amplitude can then be put in the simple form 
\begin{equation}
\mathcal A=-2\frac{M_P^2}{t}\left(\frac{{\bar m}^2}{M_P^2}-q^2e^{\sqrt 2\frac{\bar \phi}{M_P}}\right)^2=-\frac{2\mu^4}{M_P^2\,t}    
\end{equation}
 from which it is straightforward to observe that it vanishes for $q^2e^{\sqrt 2\frac{\bar \phi}{M_P}}={{\bar m}^2}/{M_P^2}\, (\mu=0)$ and is always negative otherwise. The no-force condition is readily seen to correspond, once we revert again to geometrized units, to the black hole extremality 
 \begin{equation}
q^2e^{2\bar \phi}=2{\bar m}^2.
\end{equation}
 As such, for $\alpha =1$, the leading classical force between two particles with charge $q$ and mass \eqref{mass case alpha=1} giving an effective description of a pair of the same black holes is always attractive and vanishes only for extremal states. \\

\begin{figure}
    \centering
    \includegraphics{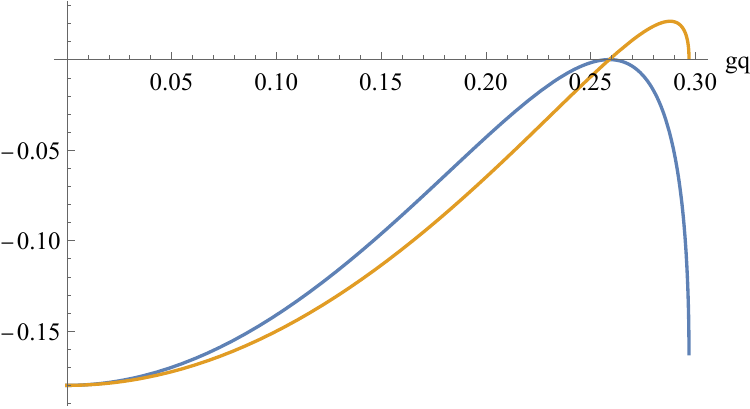}
    \caption{Comparison of the leading overall forces for the scattering of two point-like states approximating far-away black holes in blue, and one Kaluza-Klein-like state and one similar to the black hole in yellow. Here, $\alpha=0.7$, and $m=0.3$ in Planck units.}
    \label{sum forces comparaison}
\end{figure}

Turning now to generic values of $\alpha$, $\gamma(\phi)$ can be rewritten in terms of physical quantities as

\begin{equation}
\gamma(\phi)=\frac{\alpha}{1-\alpha^2}\left(1-\sqrt{1-2M_P^2(1-\alpha^2)\frac{q^2}{m^2(\phi)}e^{\sqrt 2\alpha\frac{\phi}{M_P}}}\right).
\end{equation} 
The tree-level contribution to the force between two states is given by the coefficient of the $t$-channel pole is

\begin{equation}
\mathcal A^{\mathrm{t-pole}} =4e^{\sqrt 2\alpha\frac{\bar \phi}{M_P}}q^2\bar m^2 -2\frac{\bar m^4}{M_P^2}-2\frac{\alpha^2}{(1-\alpha^2)^2}\frac{\bar m^4}{M_P^2}\left(1-\sqrt{1-2M_P^2(1-\alpha^2)\frac{q^2}{m^2_b}e^{\sqrt 2\alpha\frac{\phi}{M_P}}}\right)^2
\end{equation}
The resulting behaviour of $\mathcal A$ is represented by the blue curve in figure \ref{sum forces comparaison}. We observe again that the overall force between two particles is always attractive, even beyond the point $q^2e^{\sqrt 2\alpha\frac{\bar \phi}{M_P}}=\frac{1+\alpha^2}{2}\frac{{\bar m}^2}{M_P^2}$ corresponding to extremality, which is again found to be the only point where the force vanishes. Contrary to the Reissner-Nordstrom case, the dilatonic coupling does not allow repulsive forces beyond extremality: increasing $q$ at fixed $m$, the scalar force grows at least as strong as the gauge one. As observed in \cite{Julie:2018lfp} and later shown in \cite{Chen:2019ohv}, in the asymptotic $\phi\to\infty$ limit, the solution to \eqref{differential equation mass Julie} takes the form $\sqrt{{(1+\alpha^2)}/{2}}\left({m(\phi)}/{M_P}\right)=qe^{\frac{\alpha}{\sqrt 2}\frac{\phi}{M_P}}$. We note here that it coincides with the extremal relation.\\

We now discuss the force felt by a particle of charge $q$ and Kaluza-Klein-like mass $m(\phi)=m_Ae^{\alpha\phi}$ (in geometrized units). Our result for the effective potential for the motion of such a particle in the asymptotically flat background  metric, when applied for this case, agrees with \cite{Shiromizu:1999bm}. We will address here the attractive or repulsive nature of the leading interaction, the ${1}/{r}$ contribution, for the case of our point-like states.

The effective potential takes the form \eqref{effective potential motion in GHS metric}, with $m'=\alpha m$. If we choose the two, particle and black hole, states such that they share the same mass ($m(\bar \phi)=M$) and charge ($q=Q$), we find that, for $\alpha<1$, the overall force, proportional to $M^2+\alpha MD-e^{2\alpha \bar{\phi}}Q^2$, as shown by the yellow curve in figure \ref{sum forces comparaison}, vanishes at two points:
\begin{itemize}
    \item the point defining the extremality condition
    \begin{equation}
     M^2=\frac {e^{2\alpha\bar{\phi}}Q^2}{1+\alpha^2}  
    \label{extremal again}
    \end{equation} 
    \item and in the point 
    \begin{equation}
     M^2=(1-\alpha^2)e^{2\alpha\bar{\phi}}Q^2   
    \end{equation} 
    where the metric is on the verge of becoming complex. 
\end{itemize} 
This remains valid as long as ${m}/{q}={M}/{Q}$.\\

Denoting the point-like state approximating a black hole as $S$, we now turn to the computation of the amplitude $S\to S\phi\phi$ for the emission of a pair of dilatons due to the couplings in \eqref{mass expansion dilaton}. It takes the form

\begin{equation}
    \mathcal A(S\to S\phi\phi)=-2i\frac{\bar m^4}{M_P^2}{\bar \gamma}^2\left(\frac{1}{t-\bar m^2}+\frac{1}{u-\bar m^2}\right)-i\frac{\bar m^2}{M_P^2}\left(2{\bar \gamma}^2+{\bar \beta}\right),
\end{equation}
where ${\bar \beta}$ is now written in physical units as the other quantities. At threshold, $t,u=-\bar m^2$, and this simplifies to 

\begin{equation}
\mathcal A(S\to S\phi\phi)=-i\frac{\bar m^2}{M_P^2}{\bar \beta}=-2i\frac{\alpha^2}{1-\alpha^2}q^2e^{\sqrt{2}\alpha\frac{\bar \phi}{M_P}} \left(1-\frac{\alpha^2}{\sqrt{1-(1-\alpha^2)2M_P^2\frac{q^2}{{\bar m}^2}e^{\sqrt 2\alpha\frac{\bar \phi}{M_P}}}}\right).   
\label{tired amplitude}
\end{equation}
For any finite value $\alpha\ge1$, the amplitude is finite as well and reduces to $\mathcal A=-2iq^2e^{\sqrt 2\alpha\frac{\bar \phi}{M_P}}$ for $\alpha=1$. However, for $\alpha<1$ the amplitude  diverges at 
\begin{equation}
q^2e^{\sqrt{2}\alpha\frac{\phi}{M_P}}=\frac{1}{2(1-\alpha^2)}\frac{{\bar m}^2}{M_P^2},
\end{equation} 
corresponding to the largest charge before the metric becomes complex. It is also easily verified that the amplitude \eqref{tired amplitude} vanishes when the no force (extremality) condition \eqref{extremal again} is met.\\







\section{Summary and conclusions}
We give here an overview of the results obtained above for the existence of horizons in the black hole solution \eqref{metric alpha generic dS}. We also attempt to infer from them new bounds for the Dilatonic Weak Gravity Conjecture (DWGC). We will keep separate the discussions about asymptotically flat,  AdS and dS space-time. For AdS, this extends previous partial results for the dilatonic black hole solution given in \cite{Elvang:2007ba,Goto:2018iay}. Also, the authors of \cite{Goto:2018iay} focused on the region $r_+ \gg r_-$. Whereas, our analysis of the horizons in the dS solution is to our knowledge new. Finally, for the WGC in this cases, we follow  \cite{Antoniadis:2020xso}, where the dS-WGC bound was conjectured to be set by the boundary, in the $(QH,MH)$ plane, between the black hole solution region exhibiting both an event and a cosmological horizon and the naked singularity region with only a cosmological horizon. The limit where the cosmological and event horizons coincide corresponds to Nariai Black Hole cases. The dilaton profile is then constant $\phi= \phi_N$. Given a generic potential $V(\phi)$, a necessary condition for the Nariai black hole to exist is given by the bound \cite{Bousso:1996pn,Montero:2020rpl}
\begin{equation}
  \left|\frac{V'}{V}(\phi_N)\right|\leq 2\alpha. 
   \label{NariaiBM} 
\end{equation}
This inequality was used in \cite{Montero:2020rpl} to derive a generic bound on the potential, without a prior condition on $\phi$. For the class of potentials $V(\phi)$ given by \eqref{dilaton potential}, it easy to find generic ranges of dilaton values where the condition \eqref{NariaiBM} is not verified. However, in the different cases we discuss here, the Nariai black hole solutions are present for specific values of the dilaton field:
\begin{equation}
e^{2\alpha \phi_N}=e^{2\alpha \phi_0}\left(1-\frac{r_-}{r_c}\right)^{\frac{2\alpha^2}{1+\alpha^2}}\, .
   \label{Nariai} 
\end{equation}
and, for such values of $\phi_N$, the bound \eqref{NariaiBM} is always satisfied.
\\

We are particularly interested in the behaviour of the black hole solutions as function of the dilaton coupling $\alpha$. We notice that the $\alpha\to \infty$ and $\alpha\to 0$ limits of such black holes are approximated by Schwarzschild and Reissner-Nordstr\"om black holes, and we find a value of $\alpha= 1, 1/\sqrt{3}$ which separates between the two behaviours in flat and (A)dS backgrounds, respectively. 
\\

\underline{ {\bf The flat space-time BH and DWGC:}}
\\

We have retrieved the well-known result \cite{Gibbons:1987ps,Garfinkle:1990qj}
    \begin{equation}
     Q^2e^{2\alpha \phi_0}\le(1+\alpha^2)M^2
    \end{equation} 
\, \, \, As formulated by \cite{Heidenreich:2015nta}, we assume that the WGC corresponds to the opposite inequality.
\\

{\underline {\bf The asymptotically AdS BH and AdS-DWGC:}}
\\

\begin{itemize}
    \item  For $\alpha>{1}/{\sqrt{3}}$:
    
    The black hole solutions exhibit only one (event) horizon. It is located outside the singular surface as long as 
    \begin{equation}
     Q^2e^{2\alpha \phi_0} < (1+\alpha^2)M^2   
    \end{equation}
   and the two surfaces coincide when the inequality turns to equality. The DWGC condition is the same as in the asymptotically flat-space one.\\

    \item When $\alpha={1}/{\sqrt{3}}$:
    
    The black hole solutions still possess only one horizon. The coincidence of that horizon with the singularity is no more obtained for $Q^2e^{2\alpha \phi_0}=(1+\alpha^2)M^2$ but for a smaller charge now, saturating the inequality 
   \begin{align}
    &\frac{\left(9\left(\hat M+\sqrt{\hat M^2-\frac{2}{3}\hat Q^2}\right)+\sqrt 3\sqrt{4+27\left(\hat M+\sqrt{\hat M^2-\frac{2}{3}\hat Q^2}\right)}\right)^{1/3}}{2^{1/3}\,3^{2/3}} \nonumber \\
    &-\frac{\left(\frac{2}{3}\right)^{1/3}}{\left(9\left(\hat M+\sqrt{\hat M^2-\frac{2}{3}\hat Q^2}\right)+\sqrt 3\sqrt{4+27\left(\hat M+\sqrt{\hat M^2-\frac{2}{3}\hat Q^2}\right)}\right)^{1/3}} \ge \frac{4\hat Q^2}{3\left(M+\sqrt{M^2-\frac{2}{3}\hat Q^2}\right)}.
    \end{align}
   The expansion for small $H$ (large $L=\frac{1}{H}$, with $L$ the AdS length scale) gives
   
    \begin{equation}
    Q^2e^{(2/\sqrt 3)\phi_0}\le\frac{4}{3}M^2-\frac{4^3}{3^4}M^4H^2+o(H^2).   
    \end{equation}
    which reproduces the flat space-time case for $H\to 0$. \\
    
    \item For $0<\alpha<{1}/{\sqrt{3}}$:
    
    The black holes have both an inner and an outer horizon. The extremality condition is now obtained when the two horizons coincide and is expressed as
    
    \begin{equation}
    F_{AdS}(r_{0-})=0  
    \end{equation}
    where $F_{AdS}$ is defined by \eqref{FAdS} and $r_{0-}$ by \eqref{rzeroplus and rzerominus}. Here the two coincident horizons are located outside the singularity. The expansion for small $H$ gives 
   \begin{equation}
   Q^2e^{2\alpha\phi_0}=(1+\alpha^2)M^2+\alpha^2(1+\alpha^2)^{\frac{2}{1-\alpha^2}}cM^{\frac{3-\alpha^2}{1-\alpha^2}}H^{\frac{1+\alpha^2}{1-\alpha^2}}+o(H^{\frac{1+\alpha^2}{1-\alpha^2}}),    
   \end{equation}
  where $c$ is defined in \eqref{c-def}. From this expansion one can see that the condition tends to the flat space one in the limit $H\to 0$. Black hole states solve $F_{AdS}(r_{0-})<0$.
\\
    
    \item For $\alpha=0$:
    
 This is the well studied case of charged AdS without dilaton (see for example \cite{Nakayama:2015hga,Li:2015rfa,Harlow:2015lma,Benjamin:2016fhe,Heidenreich:2016aqi,Montero:2016tif,Conlon:2018vov,Montero:2018fns,Alday:2019qrf,Cremonini:2019wdk,Agarwal:2019crm}).\\
 
\end{itemize}

A WGC bound can be identified as the requirement of the presence of a state with a charge $Q$ and a mass $M$ that verifies an inequality opposite to the ones above respected by black holes with a horizon.\\

   In \cite{Crisford:2017zpi}, the classical decay of (near)-extremal solutions through the charged superradiance mechanism was used to obtain a WGC bound in asymptotically AdS space-time with a dilaton. The conjecture requires the existence of a state with mass $m$ and charge $q$ solving
   
   \begin{equation}
   \label{WGC superradiance}
      q\ge \frac{\Delta}{\mu}, \qquad \mathrm{with} \qquad \Delta=\frac{3H}{2}+\sqrt{\frac{9H^2}{4}+m^2}, \qquad \mu=\frac{Q}{r_+}.
   \end{equation}
   where $\Delta$ is the minimum frequency of a scalar perturbation in AdS and $\mu$ the difference between the component $A_t$ of the gauge field at infinity and on the horizon for extremal solutions.
   The condition for the onset of superradiance was obtained considering the leading order in the horizon radius for small black hole and the equation $Q^2e^{2\alpha \phi_0}=(1+\alpha^2)M^2$ to define extremal states. With these assumptions, eq. \eqref{WGC superradiance} reads
   
   \begin{equation}
   \label{WGC superradiance explicit}
    q\ge \Delta \sqrt{1+\alpha^2}.   
   \end{equation} 
  Here, for $\alpha\le1/\sqrt{3}$, we found that the extremality condition receives corrections. Accordingly, the bound from \eqref{WGC superradiance} becomes
   
   \begin{equation}
   \label{boundsuperradiance corrected}
     q\ge \frac{r_{0-}}{\sqrt{r_+r_-}}\Delta\sqrt{1+\alpha^2},  
   \end{equation}
where $r_+$ and $r_-$ are related by the condition $F_{AdS}(r_{0-})=0$,  $r_{0-}=r_{0-}(r_+,r_-)$. Again, $F_{AdS}$ is defined by \eqref{FAdS} and $r_{0-}$ in \eqref{rzeroplus and rzerominus}. For $r_+\ge r_-$, and thus for extremal solutions, $r_{0-}<\sqrt{r_+r_-}$, so that the bound \eqref{boundsuperradiance corrected} is weaker than \eqref{WGC superradiance explicit}.   
Using \eqref{rzeroplus and rzerominus} and the expansion of the extremality condition for small $H$ \eqref{expansion r- ads}, this gives at leading order:

\begin{equation}
q\ge   \Delta\sqrt{1+\alpha^2}\left(1+\frac{\alpha^2}{1-\alpha^2}cr_+^{\gamma}H^{\gamma}+o(r_+^{\gamma}H^{\gamma}) \right), 
\end{equation}
where $\gamma={(1+\alpha^2)}/{(1-\alpha^2)}$ and $c$ is a constant solution of the equation given in \eqref{c-def}. Note that the expression of the minimum frequency in AdS might also receive corrections which we expect to be sub-leading (for RN-AdS, we have $\omega=H\Delta+o(r_hH^2)$). \\

\underline{{\bf The asymptotically dS-BH and dS-DWGC}}\\

\begin{itemize}
    \item For $\alpha>{1}/{\sqrt{3}}$, the condition for the event horizon to coincide with the singularity is given by $Q^2e^{2\alpha \phi_0}=(1+\alpha^2)M^2$. It is again the same as in both asymptotically flat and  AdS space.\\

    \item When $\alpha={1}/{\sqrt{3}}$, the extremal solution solutions solve
    \begin{align}
    &\frac{\left(\frac{2}{3}\right)^{1/3}e^{-i\pi/3}}{\left(-9\left(\hat M+\sqrt{\hat M^2-\frac{2}{3}\hat Q^2}\right)+\sqrt 3\sqrt{-4+27\left(\hat M+\sqrt{\hat M^2-\frac{2}{3}\hat Q^2}\right)}\right)^{1/3}}   \nonumber\\
    &+\frac{e^{i\pi/3}\left(-9\left(\hat M+\sqrt{\hat M^2-\frac{2}{3}\hat Q^2}\right)+\sqrt 3\sqrt{-4+27\left(\hat M+\sqrt{\hat M^2-\frac{2}{3}\hat Q^2}\right)}\right)^{1/3}}{2^{1/3}\,3^{2/3}} \nonumber \\
    &=-\frac{4\hat Q^2}{3\left(\hat M+\sqrt{\hat M^2-\frac{2}{3}\hat Q^2}\right)}.   
    \end{align}
    This condition can be expanded for small $H$ as 
    \begin{equation}
    Q^2e^{(2/\sqrt 3)\phi_0}=\frac{4}{3}M^2+\frac{4^3}{3^4}M^4H^2+\mathcal{O}(M^6H^4),    
    \end{equation}
    showing that it allows for slightly greater charges than the corresponding flat space limit. It goes to the flat space-time condition for $H\to 0$.\\

    \item For $0<\alpha<{1}/{\sqrt{3}}:$
    
    For a real valued metric, one has always either two horizons or, trivially for huge masses, a naked singularity. We have found no region of parameters with a (real) metric exhibiting only a naked singularity with a cosmological horizon. 
    
    In both limits $\alpha\to 0$ and $\alpha\to {1}/{\sqrt{3}}$ limits, we retrieve the $\alpha = 0$ and $\alpha={1}/{\sqrt{3}}$ results, respectively, where a transition from a black hole to a naked singularity with cosmological horizon happens for $F(r_{0-})=0$ (defined by \eqref{polynome for alpha2>1/3} and \eqref{rzeroplus and rzerominus}). 
    
    We discuss further this case below. 
    
    \item For $\alpha=0$ 
    
    The phases of the dS-RN black hole metric are described in details in \cite{Romans:1991nq,Antoniadis:2020xso}. The condition for the existence of the black hole is
\begin{equation}
\label{flatlimit}
Q^2 \le M^2 + M^4 H^2+ {\cal O}(M^6H^4) 
\end{equation}
with $\quad M^2 H^2 \leq \displaystyle\frac{2}{27}$.
    \end{itemize}

The appearance of a complex valued metric leaves the lower bound on the black hole mass  for the $0<\alpha<{1}/{\sqrt{3}}$ asymptotically dS solutions undefined. Because previous literature focused on the asymptotically flat metric which always shows a naked singularity before reaching the complex valued metric region, it was not investigated. Here, however, the condition for the metric becoming complex, 
\begin{equation}
\label{complexlimit}
Q^2e^{2\alpha\phi_0} > \frac{M^2}{1-\alpha^2}  
\end{equation}
 can be reached within the domain of the black hole solution region and represents a new bound. \\  
 
To investigate the nature of this bound, we studied, in the asymptotically flat space, the behaviour of the interaction between point-like particles approximating at long distances the charged dilatonic black holes. The main difficulty is then to express the coupling of the dilaton to the point-like particles as a function of the charge and mass which is non-trivial for the generic solution. A first approximation was to adopt the relation between scalar and gauge charges and mass as it appears for the black hole solution at infinity. We found then that the self interaction between two such states is always attractive, even in the super-extremal case, and is null only for extremal solutions with $Q^2e^{2\alpha\phi_0}=(1+\alpha^2)M^2$. Increasing further the charge to mass ratio, the production of dilaton pairs seems to diverge for $\alpha <1$ when we reach $Q^2e^{2\alpha \phi_0}=M^2/(1-\alpha^2)$ above which the metric becomes complex. Considering instead a particle  with a mass $m(\phi)=m_Ae^{\alpha \phi}$, interacting with a black hole of same mass and charge at large distance, we found that for $0<\alpha <1$ the leading order forces between these two states cancel at the extremality but also at the point where the metric is on the verge of becoming complex. We might consider that \eqref{complexlimit} represents a new dilatonic de Sitter WGC bound for this domain of dilaton couplings ($0<\alpha <1/\sqrt{3}$), but further investigation is needed to confirm or infirm this. Also, it will be interesting to consider the effect of higher derivative corrections to the black hole solutions.



\noindent
\section*{Acknowledgments}
K.B. acknowledges the support of  the Agence Nationale de Recherche under grant ANR-15-CE31-0002 ``HiggsAutomator''.


\providecommand{\href}[2]{#2}\begingroup\raggedright\endgroup

\end{document}